\documentclass[reprint,amssymb,amsmath,aip,cha]{revtex4-1}
\usepackage{graphicx}
\usepackage[caption = false]{subfig}
\usepackage{color}
\usepackage{epsfig}
\usepackage{ifpdf}
\usepackage{bm}
\usepackage[colorlinks=true,linkcolor=blue]{hyperref}%
\expandafter\ifx\csname package@font\endcsname\relax\else
 \expandafter\expandafter
 \expandafter\usepackage
 \expandafter\expandafter
 \expandafter{\csname package@font\endcsname}%
\fi
\usepackage{cleveref}
\begin{document}
\title{Perspective: The dusty plasma experiments a learning tool for physics graduate students}
\author{Mangilal Choudhary}
\email{jaiijichoudhary@gmail.com} 
\affiliation{Institute of Advanced Research, The University for Innovation, Koba, Gandhinagar, 382426, India}
%
\begin{abstract}
The plasma is an ionized gas that responses collectively to any external (or internal) perturbations. Introducing micron-sized solid dust grains into plasma makes it more interesting. The solid grains acquire large negative charges on their surface and exhibits collective behavior similar to the ambient plasma medium. Some remarkable features of the charged dust grain medium (dusty plasma) allow us to use it as a model system to understand some complex phenomena at a microscopic level. In this perspective paper, the author highlights the role of dusty plasma experiments as a learning tool at undergraduate and post-graduate physics programs. The students could have great opportunities to understand some basic physical phenomena as well as to learn many advanced data analysis tools and techniques by performing dusty plasma experiments. How a single dusty plasma experimental device at a physics laboratory can help undergraduate and post-graduate students in the learning process is discussed.
\end{abstract}  
\maketitle
\section{Introduction}
When the gas is subjected to a strong electric field, gas atoms get ionized and the gas phase transforms into an ionized gas phase. This ionized gas consists of equal numbers of positive  (ions) and negative charges (electrons) if the gas is completely ionized. Above a threshold density of charged species (electrons and ions), the charged particles interact via long-range Coulomb interaction and capable to exhibit the collective response to an external field similar to other phases of matter. Therefore, the ionized gas medium, named plasma, is considered as the fourth state of matter \cite{chenplasmaphysicsbook,libermanprocessingbook}. In laboratory experiments, the gas is often partially ionized therefore a large number of neutral atoms present along with the electrons and ions. What does happen if sub-micron to micron-sized solid particles are introduced into the plasma?
As these solid particles come into contact with the plasma, they acquire negative charges on their surface due to the collection of higher electron current than the ion current. The role of dust grains in plasma depends on the concentration or density of charged dust. In the case of very low dust density, well separated charged dust particles only modify the characteristics of ambient plasma and it is named plasma with impurity (or dirty plasma). In the second case where dust density is high, charged dust particles experience the long-range Coulomb interaction and exhibit the collective response to the force field. In this case, plasma is named as dusty plasma \cite{shukladustybook}.
\par 
In laboratory plasma, massive dust particles ($M_d \sim 10^{-15}$ to $10^{-11}$ Kg) acquire large negative charges the order of $10^3$-$10^5$ of electron charge \cite{barkendustcharging2,goreedustcharging1}. Therefore, the dust grain medium has some remarkable or extraordinary features to differentiate with conventional two components (electrons-ions) plasma. Firstly, a large amount of charge on the dust grain surface increases the average potential energy of the dust grain compared to its average kinetic energy. The Coulomb interaction among the nearby charged dust particles determines the phase (solid, liquid, or gas) of the dust grain medium\cite{shuklasurveyofdustyplasma,shukladustybook,thomascrystalmelt2,linidustycrystal2,thomasdustycrystal1}. Secondly, extremely small charge-to-mass ratio ($Q_d /M_d$) of dust grains leads to new plasma eigenmodes at very low frequency (1-100 Hz) \cite{daw2,daw3,merlinodawreview}. The dust dynamics at such low frequency can be visualized even with naked eyes, which allows us to study the dynamics of dusty plasma medium at microscopic level \cite{merlinodawreview,mangilalpopdaw9}. Due to these novel features of dusty plasma, it can be considered as a model system to understand various phenomena happening in the physical universe \cite{shukladustybook}.\par 
It is well known that there is a physics laboratory for the undergraduate and post-graduate physics programs in the academic institutions. The purpose of establishing a physics lab along with theoretical courses is to demonstrate the role of physics to understand physical phenomena through experiments. Students would learn to design, develop, and perform experiments to understand physics laws and naturally occurring phenomena around us. The practical work engages students to develop skills, understand the process of scientific investigation and develop their understanding of concepts of physics. In summary, practicals in physical sciences at the graduate level have a great significance in the creating scientific temper and learning process of students.\par 
The physics laboratories for undergraduate and post-graduate courses are equipped with different experimental devices. Most of the experimental devices are designed to understand some particular physical phenomena or physics law. In recent years, plasma experiments are included in the postgraduate physics laboratories in higher educational institutes around the globe with the aim to create a basic understanding of the fourth state of matter (plasma)\cite{plasmalab1,plasmalab2,plasmalab3,plasmalab4}. It has been discussed earlier that dusty plasma is created in the background of plasma medium; therefore, the establishment of dusty plasma experimental setup at the physics lab can provide more experimental opportunities for the undergraduate and postgraduate students. The low-cost dusty plasma device can also be used as an ordinary DC or RF plasma device without the dust particles.
\par
A single dusty plasma device can be used to demonstrate various basic experiments, for examples,  the study of waves and oscillations \cite{merlinodawreview}, diffraction of waves \cite{diffraction}, Crystal formation,  phase transition \cite{thomascrystalmelt2} and vortex formation \cite{mangirotationpop} etc.,  with the modification in electrode configurations, discharge conditions, and discharge configurations (DC and RF discharge). Apart from this, an external electric and magnetic field can play a significant role to perform the experiments of vortex flow and rigid body rotational motion \cite{knopkamagneticrotation,vasilievdcrotationinb,dzlievarotationstratamagnetic}. This device could also be used to understand the image analysis tools using the Matlab, Python and ImageJ software \cite{imagejsoftware} to characterize the complex flow patterns and spectral analysis of waves. In the absence of dust particles (without dust), the same device can be used to perform the basic plasma experiments\cite{plasmalab1,plasmalab2,plasmalab3,plasmalab4}. The plasma experiments help the student to learn about gas discharges and the use of simple diagnostics to characterize plasma. Since plasma is a highly non-linear system, the student can get various kinds of non-linear signals to understand the data analysis tools \cite{pankajiimeseries1}.\par
The paper is organized as follows: Section~\ref{sec:exp_setup} deals with the detailed description of the dusty plasma experimental setup. The experiments on dusty acoustic waves and opportunities for students are discussed in Sec.\ref{sec:daw_results}. In Sec.\ref{sec:diffraction_daw}, diffraction of dust acoustic waves by a cylindrical object and its application for understanding the diffraction of sound waves are discussed. The experiments on dusty plasma Crystal and phases of dust grain medium are explored in Sec.\ref{sec:crystal_dust}. The vortex formation and rotational motion of dust grain medium in the absence and presence of an external magnetic field are discussed in Sec.\ref{sec:vortex_rotation}. An opportunity to use the dusty plasma images in learning various image processing and image analysis tools is explored in Sec.\ref{sec:image_analysis}. In the absence of dust particles, plasma is a highly non-linear system. A discussion on time-series data and data analysis techniques is given in Sec.\ref{sec:data_analysis}. A brief summary of the proposed dusty plasma experiments at graduate-level physics programs along with concluding remarks is provided in Section~\ref{sec:summary}.
\section{Dusty Plasma Experimental Setup} \label{sec:exp_setup}
\begin{figure*}
 \centering
\subfloat{{\includegraphics[scale=0.80]{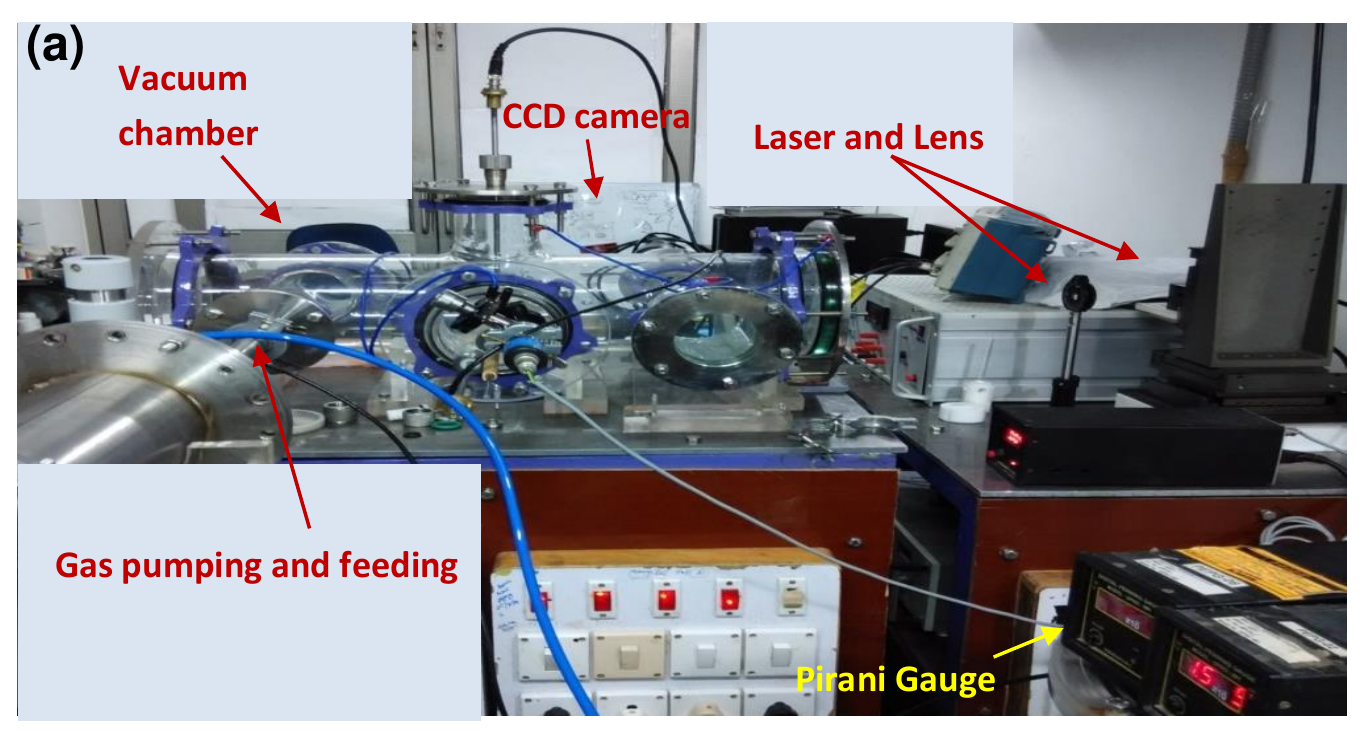}}}%
\qquad
 \subfloat{{\includegraphics[scale=0.8000]{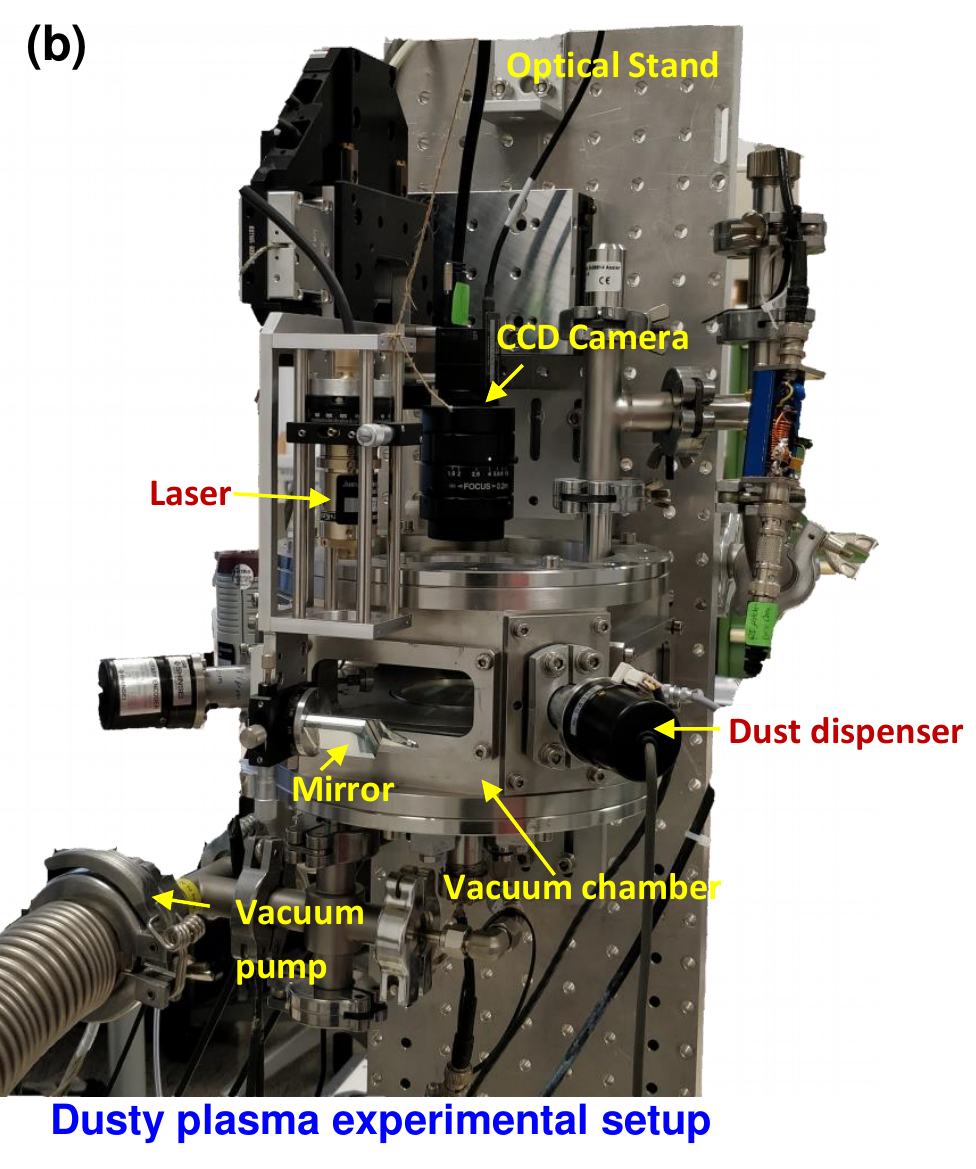}}}
 \caption{\label{fig:fig1}(a) A typical Dusty plasma device made up of glass tube. (b) A dusty plasma device made up of aluminium chamber \cite{dietzfcctobcc}. Such devices can be used to perform dusty plasma experiments in the absence or presence of an external magnetic field.} 
 \end{figure*}

A borosilicate glass tube or stainless steel (SS-304) of appropriate inner diameter (5 cm to 15 cm), thickness (8 to 14 mm), and length (10 cm to 50 cm) along with sufficient radial ports could be used for making a dusty plasma experimental device (or plasma device) \cite{mangilalrsi,Agarwalrotation}. The Axial and radial ports of the tube are used for pumping, gas feeding, holding electrodes, pressure measurement gauges, and dusty/plasma diagnostics purposes. A geometrical (3D) view of a typical dusty plasma setup made up of a glass tube is shown in Fig.\ref{fig:fig1}(a). For plasma production between two well-separated planar electrodes, either radio-frequency (RF) power source (P $\sim$ 100 W) or direct current power supply ($V_{dc}$ $>$ 600 V, $I_{dc}$ $>$ 0.5 A) are mainly used. A rotary pump attached to the glass tube or SS tube is used to create a base vacuum of $< 10^{-3}$ mbar. The relative pressure inside this vacuum chamber is measured using a Pirani gauge. A needle valve or mass flow controller (MFC) attached to the vacuum chamber is used to feed required gas into the vacuum chamber to perform experiments \cite{mangilalrsi,pdasw,Agarwaldustrot2}.
Apart from the glass or SS-304 vacuum (or experimental) chamber, an Aluminium chamber can also be used to make a dusty plasma experimental setup \cite{thomasmagnetizeddusty,mangimagneticrotation,dietzfcctobcc,chudustycrystal3} The view of the aluminum dusty plasma device is shown in Fig.\ref{fig:fig1}(b). This setup is currently used to study the magnetized dusty plasma \cite{mangimagneticrotation} at Justus-Liebig University Giessen, Germany. Such tabletop dusty plasma devices are more suitable to study the magnetized dusty plasma. It should be noted that there are different types of dusty plasma devices \cite{chudustycrystal3,hariprasaddustycrystal,merlinononlineardaw,ddwfortov,dustroataionmagnetickarasev,nakumuradustydevice} which can also be used to explore the physics at undergraduate or postgraduate level.\par
 \begin{figure*}
 \centering
\subfloat{{\includegraphics[scale=0.5500]{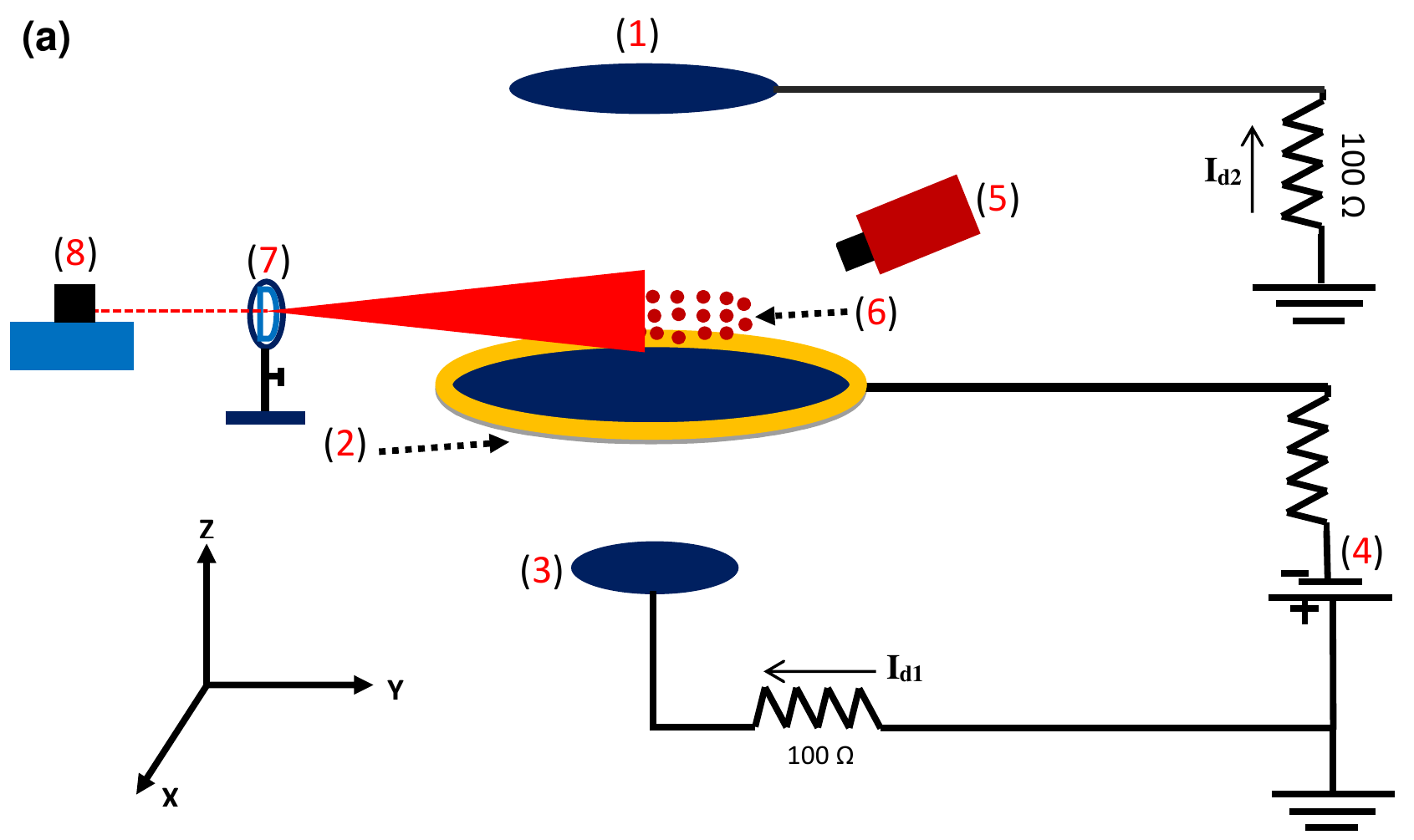}}}
 \qquad
 \subfloat{{\includegraphics[scale=0.7500]{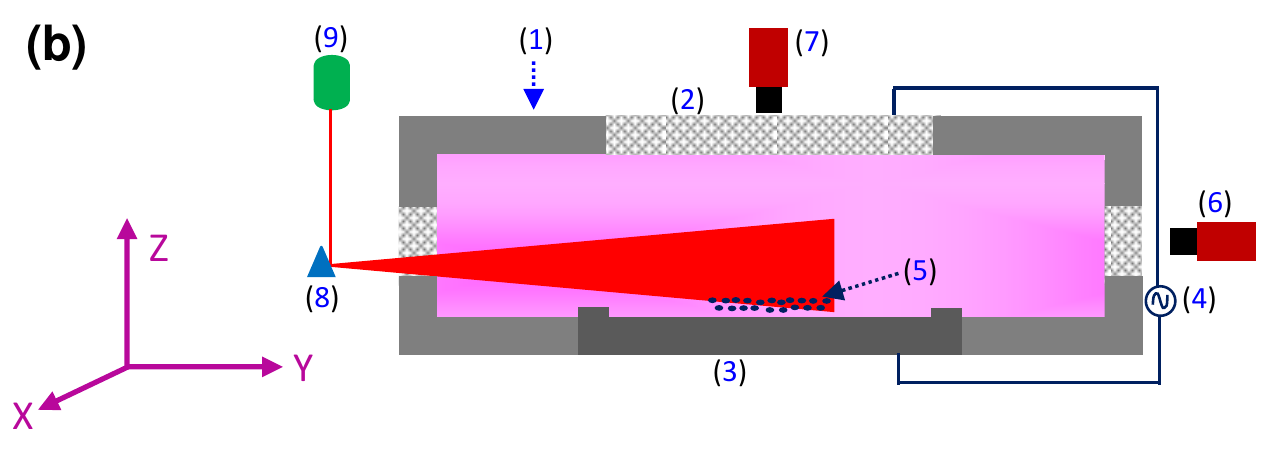}}}
 \caption{\label{fig:fig2}(a) A double anode configuration using DC power source to produce dusty plasma. (1) and (3) anodes, (2) cathode, (4) DC power supply, (5) CCD camera, (6) dust particles, (7) cylindrical lens and (8) red laser. The direction of current flowing between cathode and anodes is marked by arrows. (b) A schematic diagram of dusty plasma experiments which is operated by an RF power source. (1) Vacuum chamber, (2) upper transparent coated electrode, (3) lower metal electrode, (4) RF power source, (5) dust grains, (6) and (7) CCD cameras, (8) mirror and (9) red laser.} 
 \end{figure*}
Once the plasma is produced in the vacuum chamber, dust grains are injected into the plasma volume using a dust dispenser. The dust particle can be sub-micron to micron-sized mono-dispersive glass (plastic) or poly-dispersive particles of mass density $\sim$ 1-2 $gm/cm^3$. The dust grains in contact with plasma acquire negative charges order of $10^3$ to $10^5$ $e^{-}$ and confined near the sheath region by balancing upward forces (electric and thermophoresis forces) and downward forces (gravitational and ion drag forces) \cite{barnesdustforces,nitterdustforces1,Charging}. Here $e^{-}$ is the charge of an electron. The charged dust particles are illuminated by a combination of low power red or green laser (30 to 100 mW) and plano-convex lens. The scattered light coming from charged dust particles are captured using a high frame rate ($>$ 20 fps) and high resolution ($>$ 2 MP) CCD or CMOS camera. A typical schematic diagram of dusty plasma experiment in DC discharge configuration \cite{mangilalpop} is shown in Fig.\ref{fig:fig2}(a). A schematic diagram of dusty plasma experiments in radio-frequency \cite{dietzfcctobcc} (or DC) discharge configuration is depicted in Fig.\ref{fig:fig2}(b). The captured video or frames by CCD or CMOS camera are stored on PC for further analysis. Computer-based software such as Videomach, ImageJ along with the MATLAB, Python image processing tools is used to analyze stored image data (or frames) for further understanding the dynamics of dusty plasma medium.\cite{imagejsoftware,piv}. Sometimes different diagnostics such as Langmuir single probe \cite{probemerlino}, double probe \cite{doubleprobemalter} and emissive probe \cite{emissivesheehan} can be used to characterize the backgrounds plasma of dust grain medium. 
\section{Dust Acoustic Waves} \label{sec:daw_results}
\begin{figure*}
 \centering
\subfloat{{\includegraphics[scale=0.900]{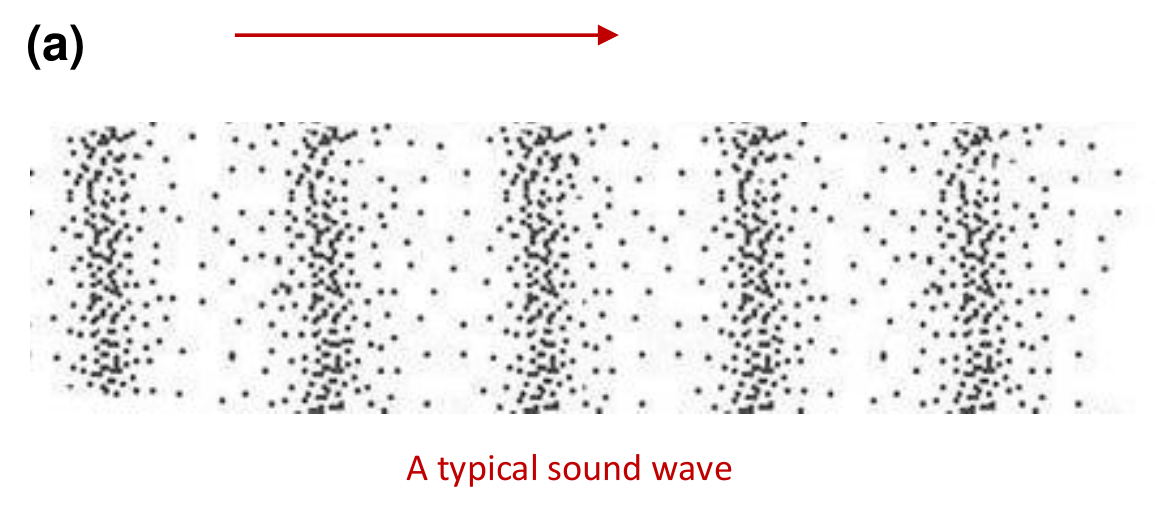}}}
\vspace*{0.15in}
 \subfloat{{\includegraphics[scale=0.600]{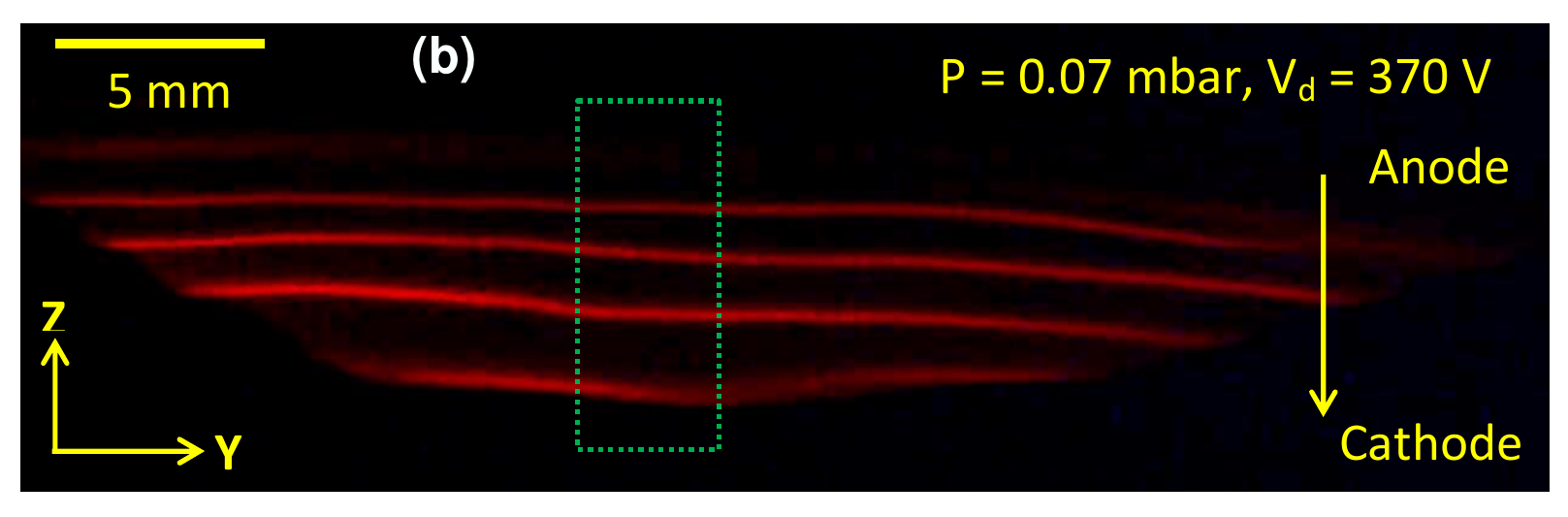}}}
 \qquad
\subfloat{{\includegraphics[scale=0.600]{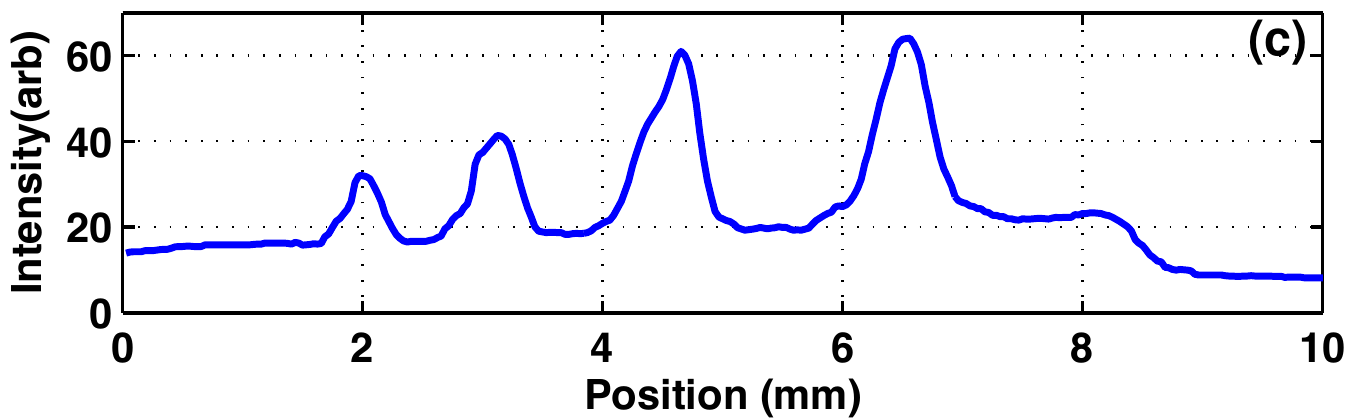}}}
 \qquad  
 \subfloat{{\includegraphics[scale=0.600]{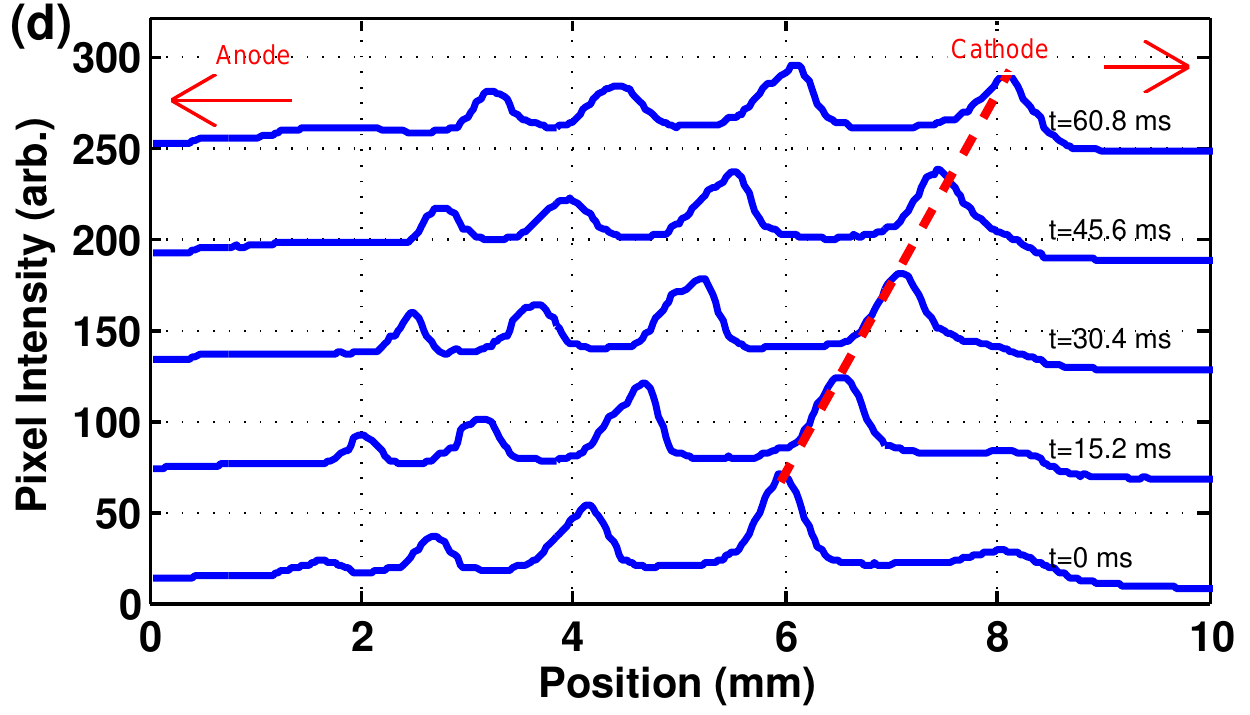}}}
 \qquad
 \subfloat{{\includegraphics[scale=0.600]{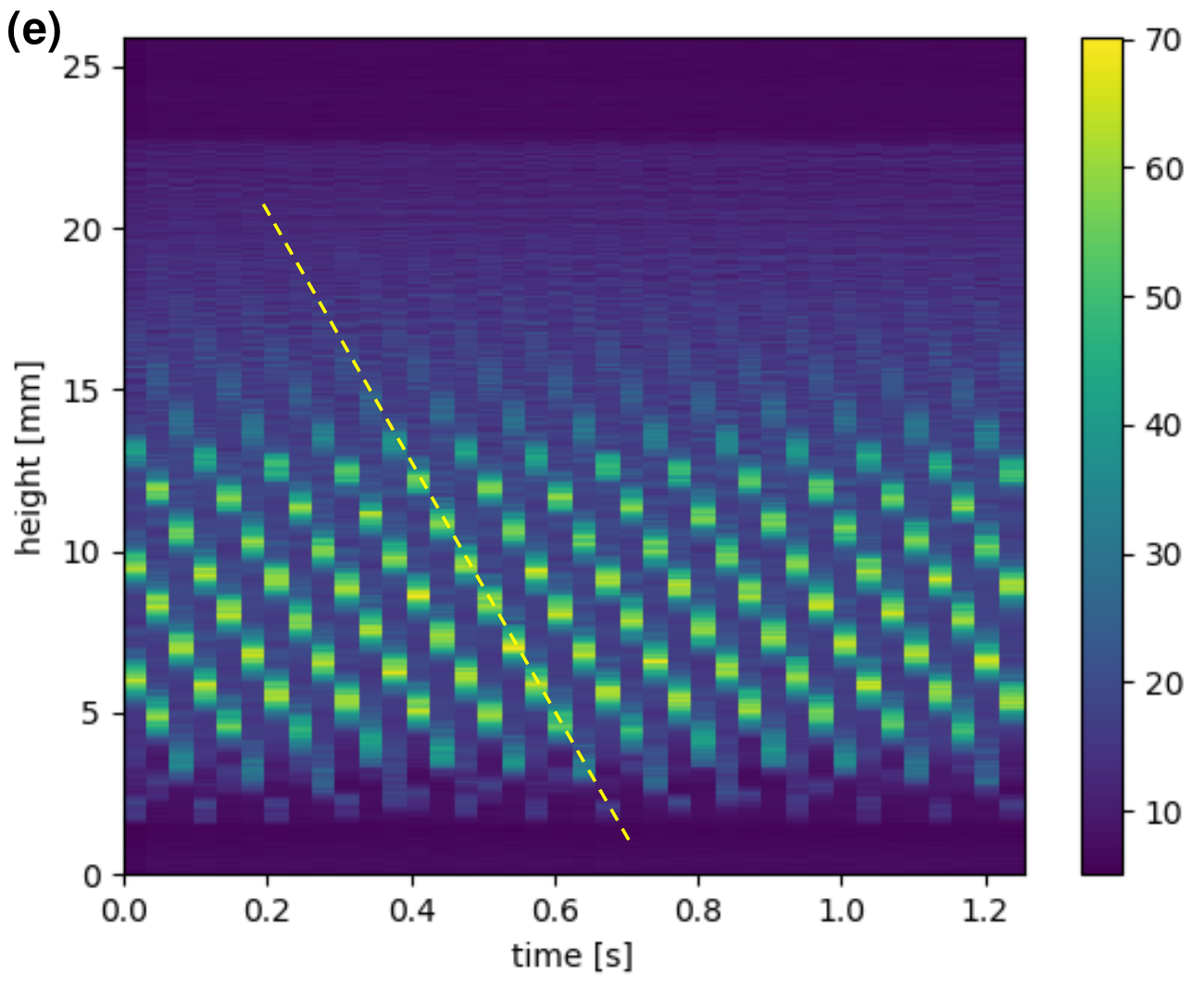}}}
 \caption{\label{fig:fig3}(a) A typical sound wave in a gas medium. (b) A snapshot of dust grain medium in Y-Z plane while dust acoustic waves are propagating. (c) An average intensity plot of selected region in Fig.\ref{fig:fig2}(b) (top to bottom).(d) An average Intensity plots of same selected region (Fig.\ref{fig:fig2}(b)) at different times. It represents propagation of dust acoustic waves from anode to cathode. Fig.\ref{fig:fig3}(b)--(d) are reproduced from [Choudhary \textit{et al.}, Phys. Plasmas 23, 083705 (2016)], with the permission of AIP Publishing. (e) A typical space-time plot which is constructed from consecutive 50 frames of DAWs using MATLAB script.}
 \end{figure*}

Wave in the gas, liquid, and the solid medium is an important topic for physics students. They are generally familiar with sound waves in different mediums which are the results of the elastic displacement of the particles of the medium about their equilibrium position. Since the motion of particles in the medium is forth and back along the direction of propagation of waves, the sound waves are the longitudinal waves. A typical sound wave in a gas medium is displayed in Fig.\ref{fig:fig3}(a). Similar to the well-known mediums, dusty plasma medium also supports very low frequency ($<$ 100 Hz) acoustic modes (DAW) \cite{daw2,daw3,merlinodawreview,mangilalpop,icpddw,mangilaldawmagneticfield}. This novel feature of dusty plasma attracts undergraduate and postgraduate students to understand the wave motion in any medium (gas, liquid, and solid) by performing experiments on dust-acoustic waves in dusty plasma device.\\
The excitation of dust-acoustic waves in dusty plasma device is possible in direct current (DC) discharge as well as in radio-frequency (RF) discharge configurations\cite{daw2,barkendaw1,dawmerlino,highlyresolvedccpdaw3,icpddw,mangilaldawmagneticfield,mangilalpop}. A schematic diagram of experimental dusty plasma setup (DC and RF discharge) to study the acoustic waves is shown in Fig.\ref{fig:fig2}. The dust acoustic waves excited in the DC discharge configurations (Fig.\ref{fig:fig2}(b)) are displayed in Fig.\ref{fig:fig3}(b). The possible causes for excitation of DAWs include ion-streaming instability and dust-acoustic instability are discussed in detail in the references \cite{ionstreaming,icpddw}. In Fig.\ref{fig:fig3}(b), bright vertical bands (or red bands) are nothing but the compressed wavefronts of the DAW, and the dark region (or low dust density region) corresponds to the rarefaction regions. Since the intensity of a bright band (wavefront) is proportional to the dust density, the intensity plots along the propagation direction at different times help to obtain the wave parameters such as wavelength ($\lambda$), phase velocity ($v_d$), and frequency ($f_d$). Intensity profile corresponding to Fig.\ref{fig:fig3}(b) is shown in Fig.\ref{fig:fig3}(c). In Fig.\ref{fig:fig3}(d), the intensity profiles of DAWs at different times are plotted. For getting intensity profile plots, one can use ImageJ (free available) software, Matlab and Python-based image processing and analysis tools, some other image analysis tools. It is also possible to get the space-time plots (see Fig.\ref{fig:fig3}(e)) using the recorded frames of propagating DAW with help of the MATLAB or Python. The space-time plots are used to characterize the dust acoustic waves \cite{melzerdustwavesinb}.\par
 \begin{figure}
\centering
 \includegraphics[scale= 0.8000]{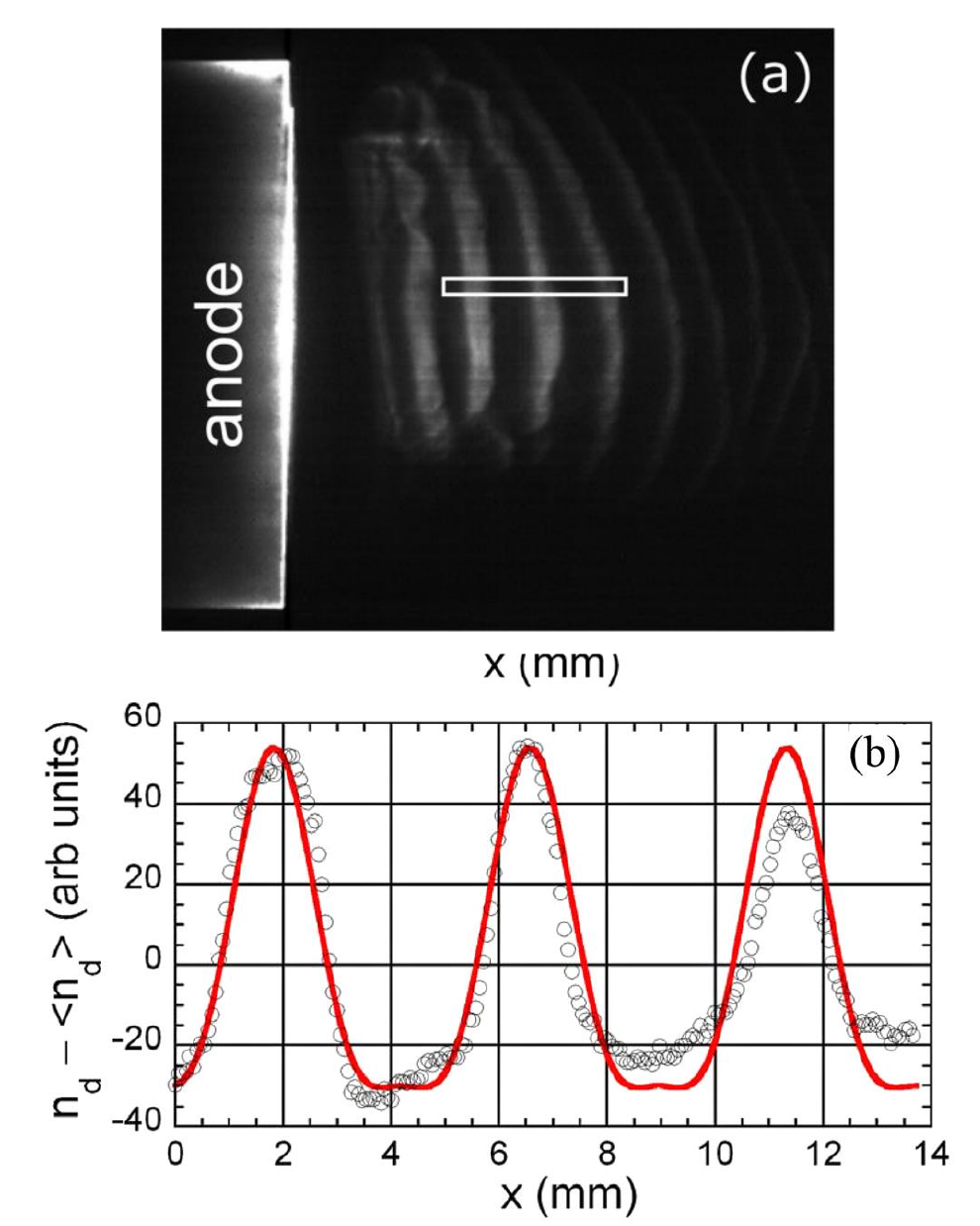}
\caption{\label{fig:fig4} (a) A snapshot of dust grain medium while waves are propagating in direction of ions flow. (b) Average intensity profile of dust acoustic wave in a selected region (rectangular) and a fitted non-linear harmonic function over the intensity profile. Fig.\ref{fig:fig4} is reproduced from [Merlino \textit{et al.}, Phys. Plasmas 19, 057301 (2012)], with the permission of AIP Publishing.}
\end{figure}
Apart from linear DAWs, various kinds of nonlinear waves can also be excited in dusty plasma using this dusty plasma device in DC discharge configuration \cite{merlinononlineardaw,pdasw}. A single video frame of dust grain medium during the propagation of nonlinear dust acoustic wave is presented in Fig.\ref{fig:fig4}(a).  In Fig.\ref{fig:fig4}(b), intensity profile of propagating nonlinear wave (Fig.\ref{fig:fig4}(a)) is plotted. These both images are taken from the original paper of Merlino \textit{et al.}\cite{merlinononlineardaw}. The nonlinear characteristics of the excited dust-acoustic wave can be verified by fitting a harmonic function (sine or cosine) of the fundamental frequency ( $f_d$)  and its harmonics to the intensity profile of propagating DAW, which is shown in Fig.\ref{fig:fig4}(b).
Space-time plots can also be used to get the frequency spectrum of propagating acoustic waves through dust grain medium\cite{tadsenspactimewave}. One can also identify the nonlinear characteristics of dust-acoustic waves based on the frequency spectrum obtained from space-time plots. There is also a possibility to modulate the self-excited dust acoustic waves and excite the linear and nonlinear waves in the dust grain medium by external forcing. In summary, it would be interesting for undergraduate and postgraduate students to explore the linear and nonlinear waves in the dust grain medium that relates the wave motion in a different medium.\\
\section{Diffraction of Dust Acoustic Waves} \label{sec:diffraction_daw}
Diffraction is an intrinsic property of the waves (mechanical waves and electromagnetic waves) in any kind of medium or vacuum. It describes the change in the direction of waves as it travels around an obstacle (barrier) or between a gap of the barrier. In daily life, we hear some noise (sound) of speaking persons from adjacent rooms through door openings which is a result of the diffraction of sound waves. The diffraction or bending of sound waves around an obstacle can be demonstrated by the study of the water waves in a ripple tank (references therein\cite{diffraction}). The amount of diffraction (spreading or bending) of the wave depends on the size of the object and the wavelength of the wave.\par
Instead of a water medium, a dusty plasma medium can be used as a model experimental system to demonstrate the diffraction of sound waves around an obstacle. The diffraction of dust acoustic waves from a cylindrical or spherical object can help to understand the diffraction phenomena of sound waves. It is possible to change the wavelength of DAW by altering the discharge parameters \cite{mangilalpop} which help to understand the amount of diffraction around different sized cylindrical objects. The dusty plasma experimental setup with a larger sized cathode ($>$ 10 cm) and smaller sized anode ($<$ 2 cm) configuration in DC discharge is suitable to explore the diffraction of DAWs.\cite{diffraction} \par
\begin{figure*}
 \centering
\subfloat{{\includegraphics[scale=0.7600]{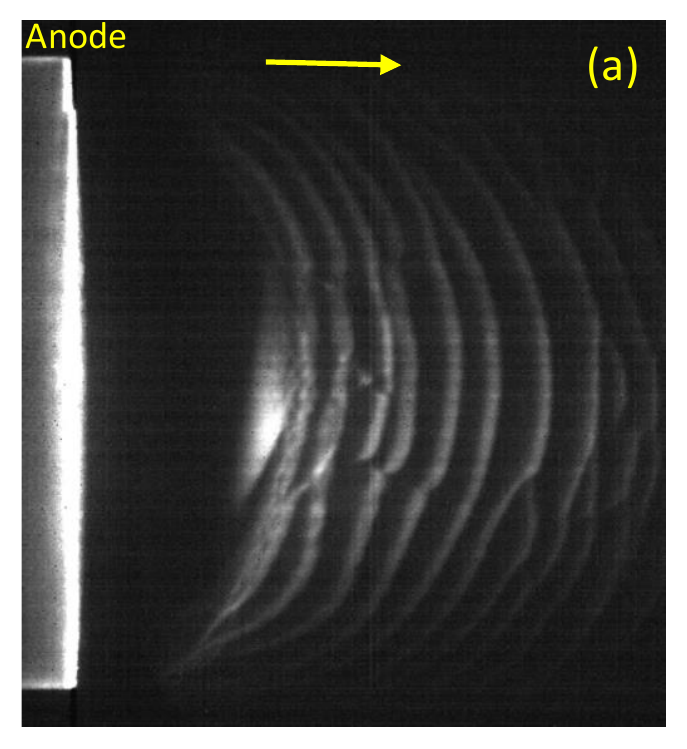}}}
\vspace*{0.25in}
\hspace*{0.45in}
 \subfloat{{\includegraphics[scale=0.76500]{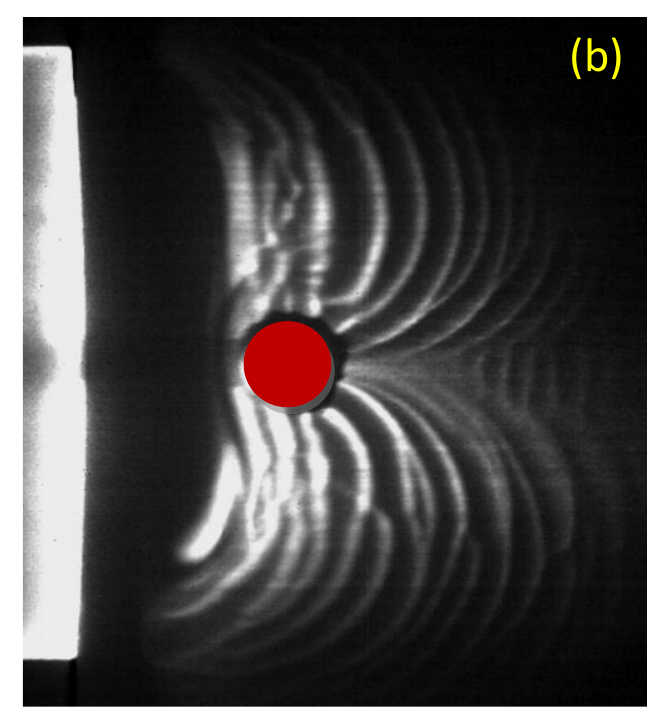}}}
 \qquad
\subfloat{{\includegraphics[scale=0.72000]{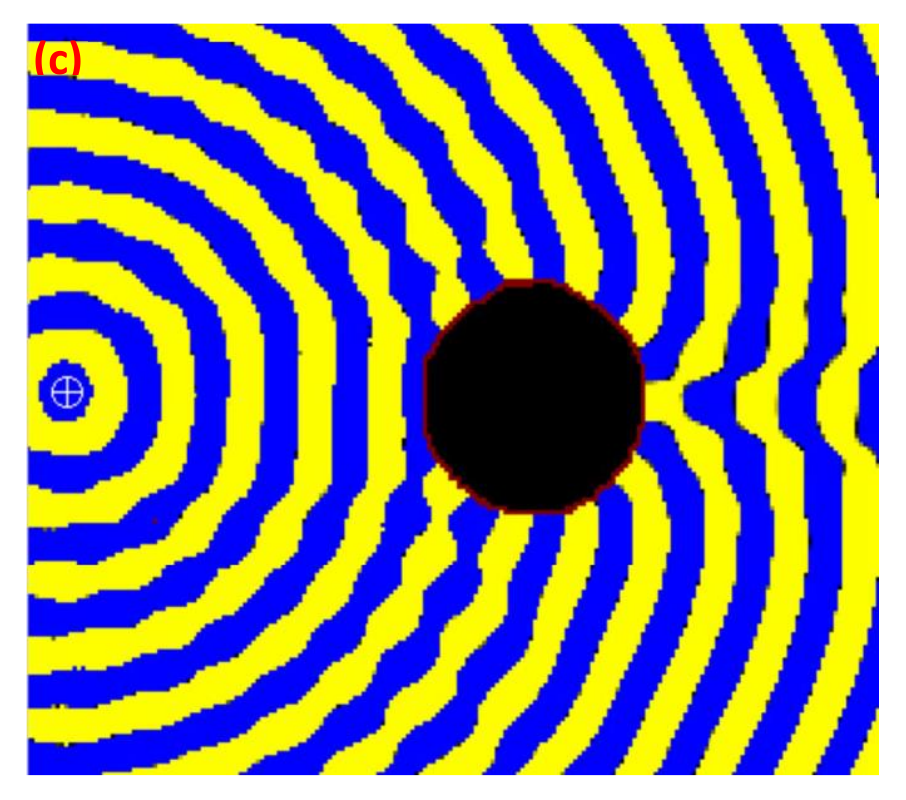}}}
 \qquad  
 \caption{\label{fig:fig5}(a) A single frame video image of dust acoustic waves in 2D plane without a cylindrical object. (b) Dust acoustic waves with the cylindrical object. A white rectangular on left side in both images is nothing but an anode. (c) Image of a ripple tank simulation of waves (originate at a point source) interacting with a circular object in 2D plane. All images of Fig.\ref{fig:fig5} are reproduced from [Kim \textit{et al.}, Phys. Plasmas 15, 090701 (2008)], with the permission of AIP Publishing.} 
 \end{figure*}
Kim \textit{et al.}\cite{diffraction} reported experimental results on the diffraction of dust-acoustic waves by a cylindrical object in weakly magnetized DC discharge plasma. In their experiments, the cathode was the grounded chamber and the anode was a 2.5 cm diameter metal disk. In Fig.\ref{fig:fig5}(a), the DAWs appear as bright vertical fringes that propagate from anode to cathode. The diffraction of DAW around a cylindrical rod (obstacle) was studied using the video images at different times\cite{diffraction}. The bending of DAW (diffraction) around the cylinder is shown in Fig.\ref{fig:fig5} (b). A ripple tank (shallow water waves in a tank) is considered a model to study the sound waves in two-dimensions (2D). Therefore, a more realistic simulation of the diffraction of sound waves in 2D can be obtained using a ripple tank. Image of a ripple tank simulation of waves (originating from a point source) interacting with a circular object is depicted in Fig.\ref{fig:fig5}(c).
Since the dust acoustic waves and sound waves obey a similar set of equations, the resulting diffraction patterns can be compared in both cases\cite{diffraction}. Thus, a dusty plasma medium can be considered as an excellent model system to learn the diffraction of sound waves at the graduate level in the physics laboratory.
\section{Crystal Formation and Phase Transition} \label{sec:crystal_dust}
In solids, atoms or molecules are closely packed. Solids could be either in crystalline or amorphous form. In crystalline solids, atoms or molecules are arranged in an ordered (long-range order) pattern whereas atoms or molecules have a short-range order in the amorphous solids. The crystalline solids can also be categorized into single-crystal solids and poly-crystalline solids. The poly-crystalline solids consist of multiple single crystal regions (grains) and the boundary separating these regions is called the grain boundary. A unit cell (smallest repeating unit) is considered as a building block of a crystalline solid. This unit cell is described by lattice vectors (lengths of each side of the unit cell) and the angle between lattice vectors. The length of lattice vectors and angles between them differentiate the types of unit cells of a crystal structure. X-ray spectroscopy is a diagnostic tool to explore the crystalline properties of solid materials. In the spectroscopic technique, one can get the diffraction patterns of scattered X-rays at different planes of crystal but difficult to realize the different crystal basis of three-dimensional (3D) crystals.\par
\begin{figure*} 
 \centering
\subfloat{{\includegraphics[scale=0.559000]{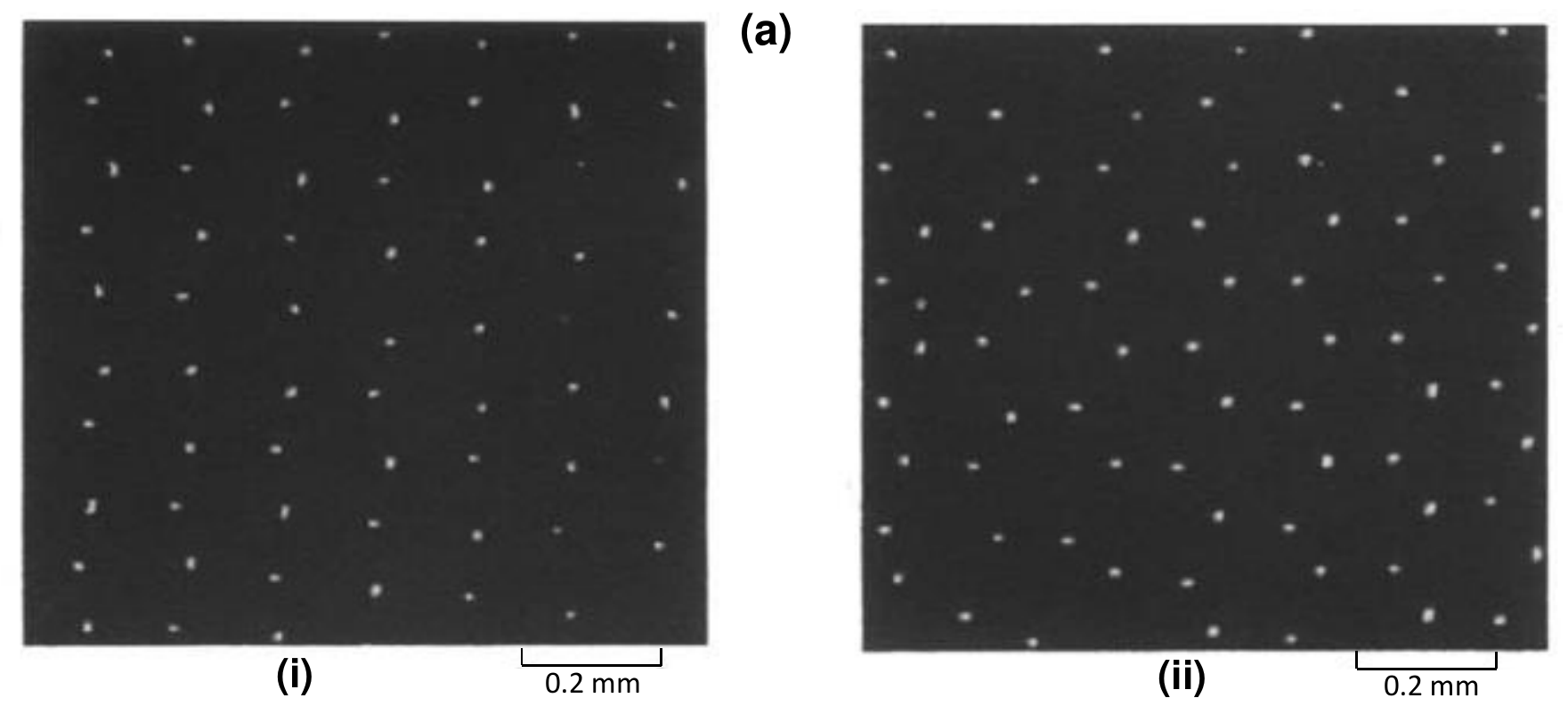}}}
\hspace*{0.45in}
 \subfloat{{\includegraphics[scale=0.5900]{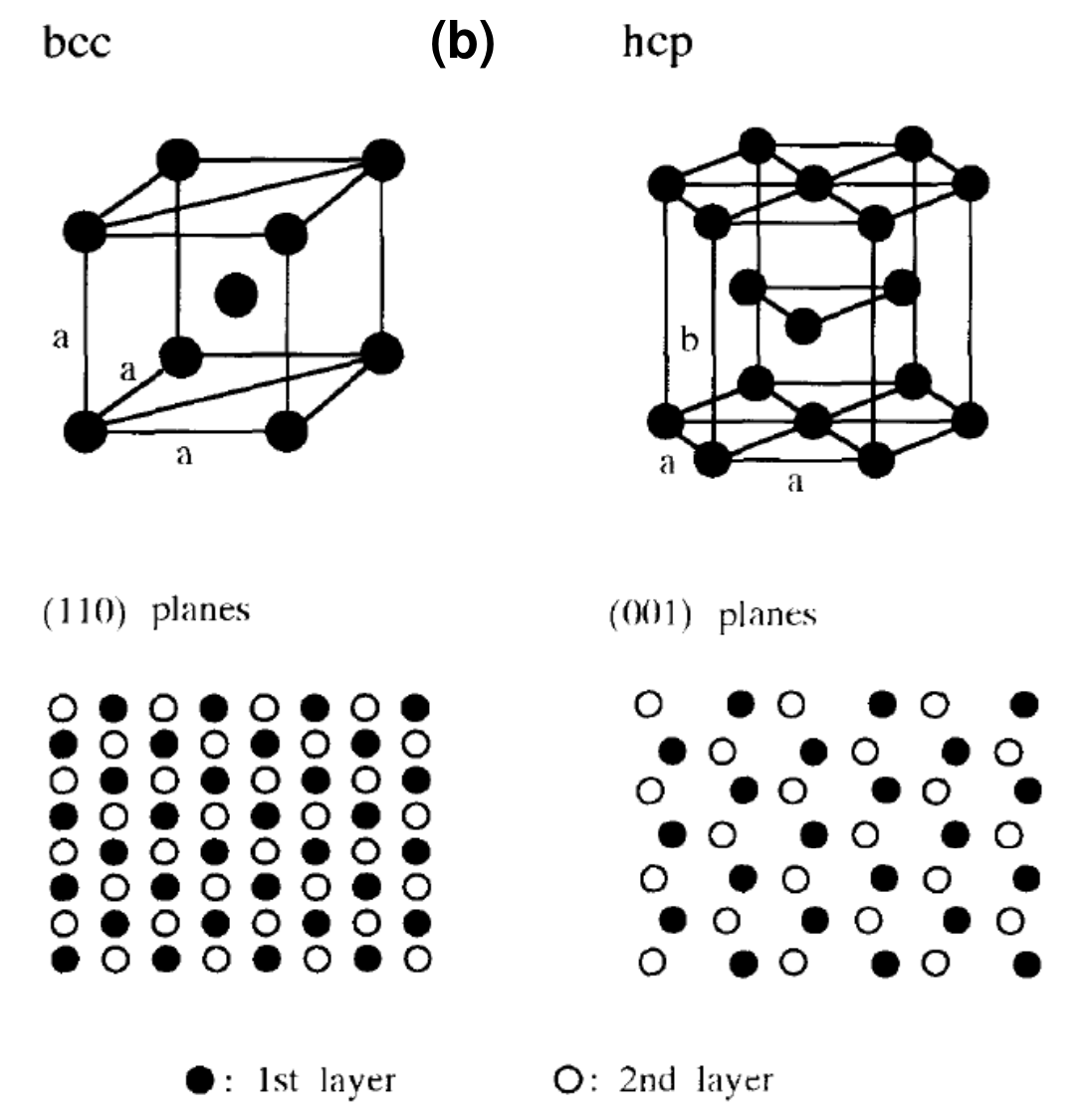}}}
\vspace*{0.2415in}
 \qquad
\subfloat{{\includegraphics[scale=0.70000]{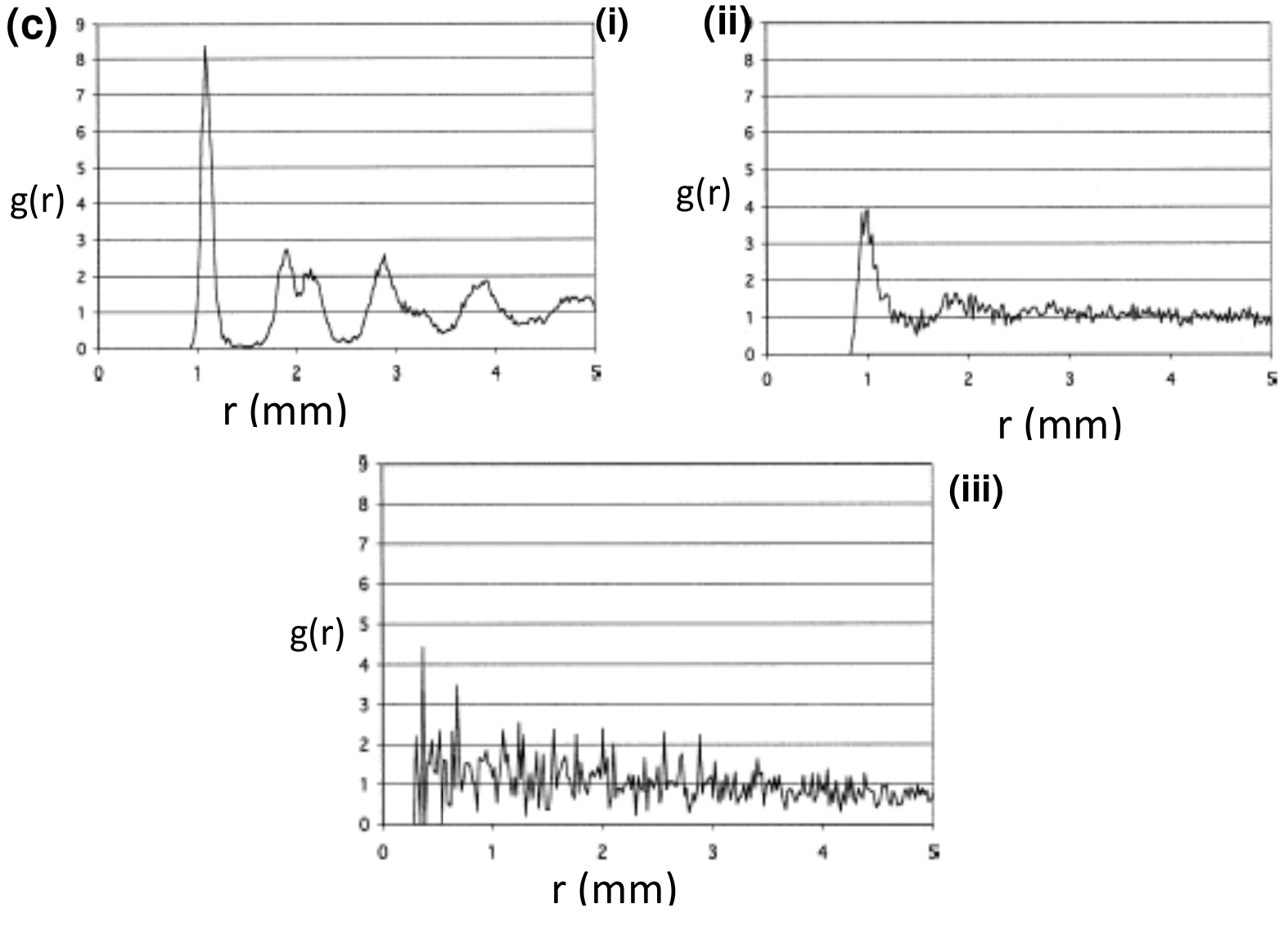}}}
 \caption{\label{fig:fig6}(a) a single video frame image of the Coulomb lattice of particles with about 10 micron diameter at power of 1 W and pressure of 200 mTorr. They show the body-centered cubic (bcc) and hexagonal close-packed (hcp). (b) Relative position of the particles in two adjacent planes in bcc (110) and hcp (001) structures. The particles in the next vertical plane are represented by filled circles while those in the original plane are shown by open circles. Images of Fig.\ref{fig:fig6}(a) and Fig.\ref{fig:fig6}(b) are Reprinted from Physica A, 205 , J. H. Chu and Lin I , Coulomb lattice in a weakly ionized colloidal plasma, 183-190, Copyright (1994), with permission from Elsevier. (c) The normalized radial correlation function for a (i) solid, (ii) liquid and (iii) gaseous state. Fig.\ref{fig:fig6}(c) is Reprinted from Advances in Space Research, 34, Smith \textit{et al.}, Dusty plasma corellation function experiment, 2379–2383, Copyright (2004), with permission from Elsevier.} 
 \end{figure*}
For understanding the crystalline structure and phase transition in solids at undergraduate and postgraduate physics programs, a realistic model system is required. In dusty plasma experiments, it is possible to create a Coulomb crystal (2D and 3D) of micron-sized charged dust particles. We term such structure a dusty plasma crystal\cite{thomascrystalmelt2,chudustycrystal3,linidustycrystal2,thomasphasetransition1,dietzfcctobcc}. One can see the dusty plasma crystal even with naked eyes and can realize the periodic arrangement of the atoms in crystalline solids as shown in Fig.\ref{fig:fig6}(a). White dots in both figures represent the dust grains. An arrangement of dust particles in a 2D plane along a vertical direction is depicted in Fig.\ref{fig:fig6}(b) that represent the types of dust crystal structures (bcc and hcp). The crystalline nature of dust grain medium is confirmed through the characteristics parameters such as Voronoi diagram \cite{hariprasaddustycrystal} and radial pair correlation function, $g(r)$\cite{paircorrelationphasesdusty,melzercrystaldustygr}.\par 
The dusty plasma experimental device either in DC or RF discharge configuration can be used to obtain the dusty plasma crystal at an appropriate discharge conditions\cite{melzercrystaldustygr,thomasdustycrystal1,hariprasaddustycrystal}. 
Using the freely available image analysis software or tools, students can analyze the dusty plasma crystal properties and correlate the results to understand the single crystalline, polycrystalline, and amorphous solids. By changing the discharge parameters such as gas pressure and input power, the melting of crystal or phase transition can be understood by obtaining the radial pair correlation function. The profile of $g(r)$ against radial distance ($r$) measures the phases of the dust grain medium \cite{paircorrelationphasesdusty,melzercrystaldustygr}. Plots of $g(r)$ in Fig.\ref{fig:fig6}(c) for different discharge conditions represent the solid, liquid, and gaseous phase of dust grain medium.
\section{Vortex and rigid rotational motion} \label{sec:vortex_rotation}
\begin{figure*}
 \centering
\subfloat{{\includegraphics[scale=0.6000]{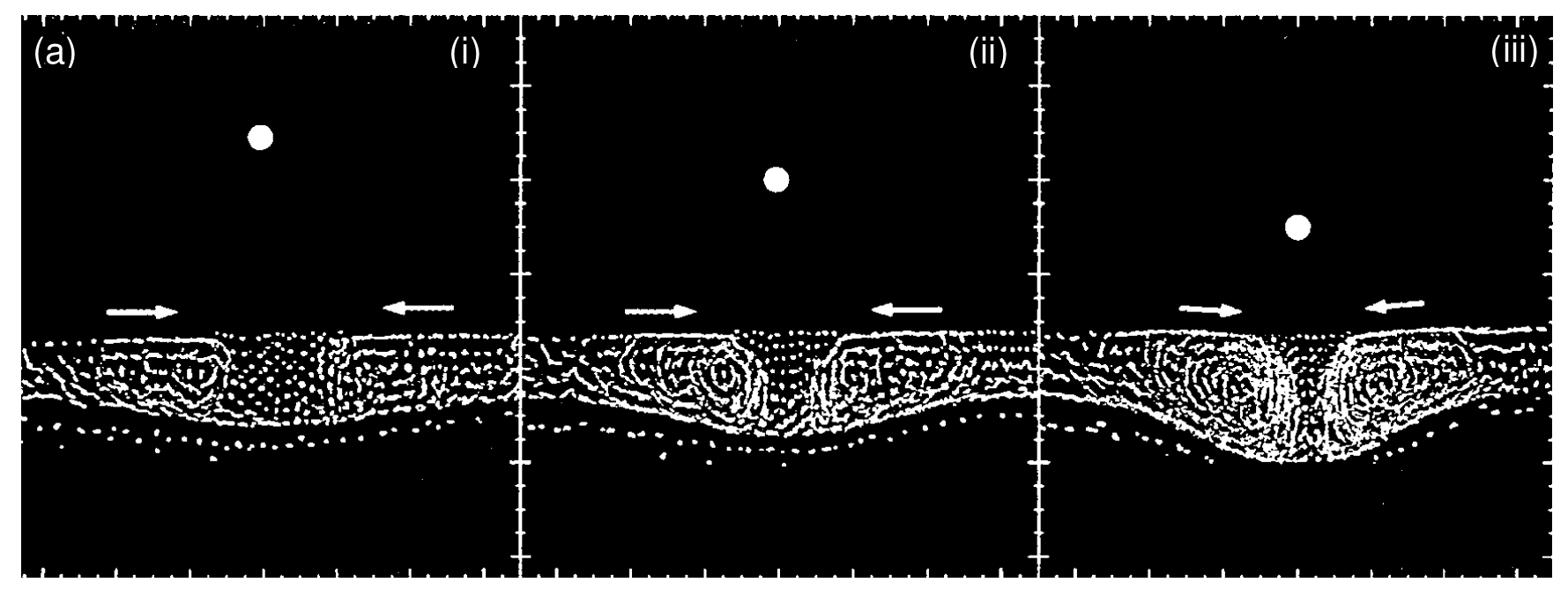}}}
\vspace*{0.2315in}
 \qquad
 \subfloat{{\includegraphics[scale=0.600]{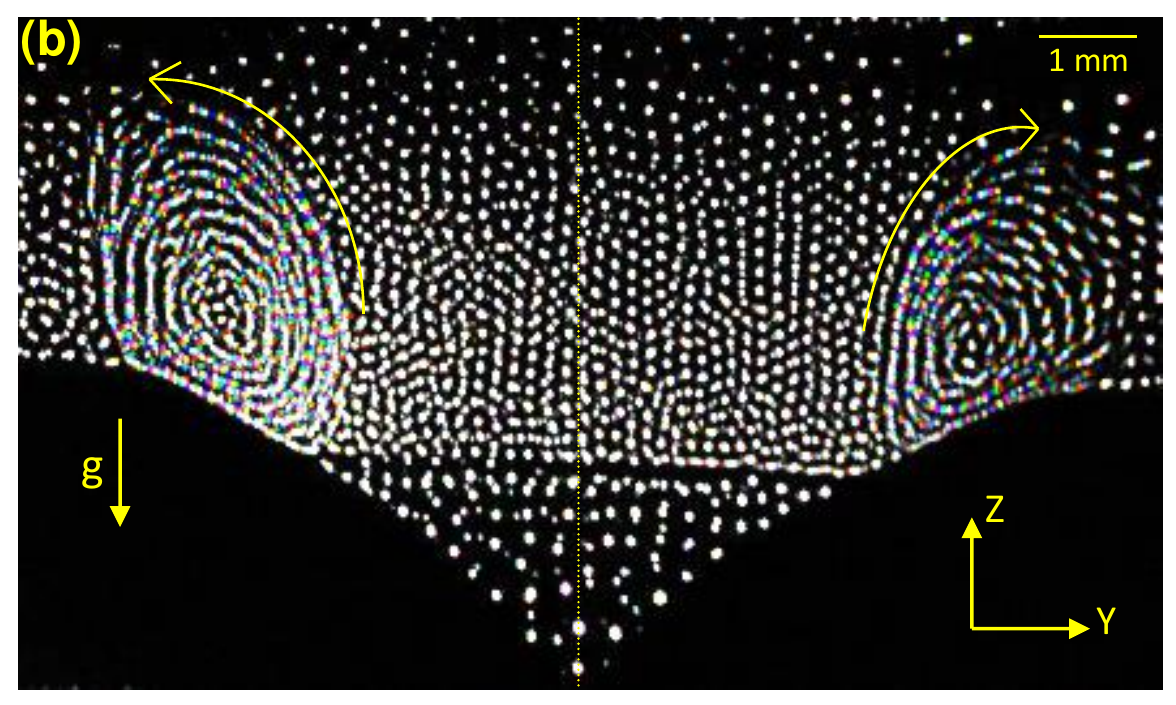}}}%
\hspace*{0.35in}
\subfloat{{\includegraphics[scale=0.6000]{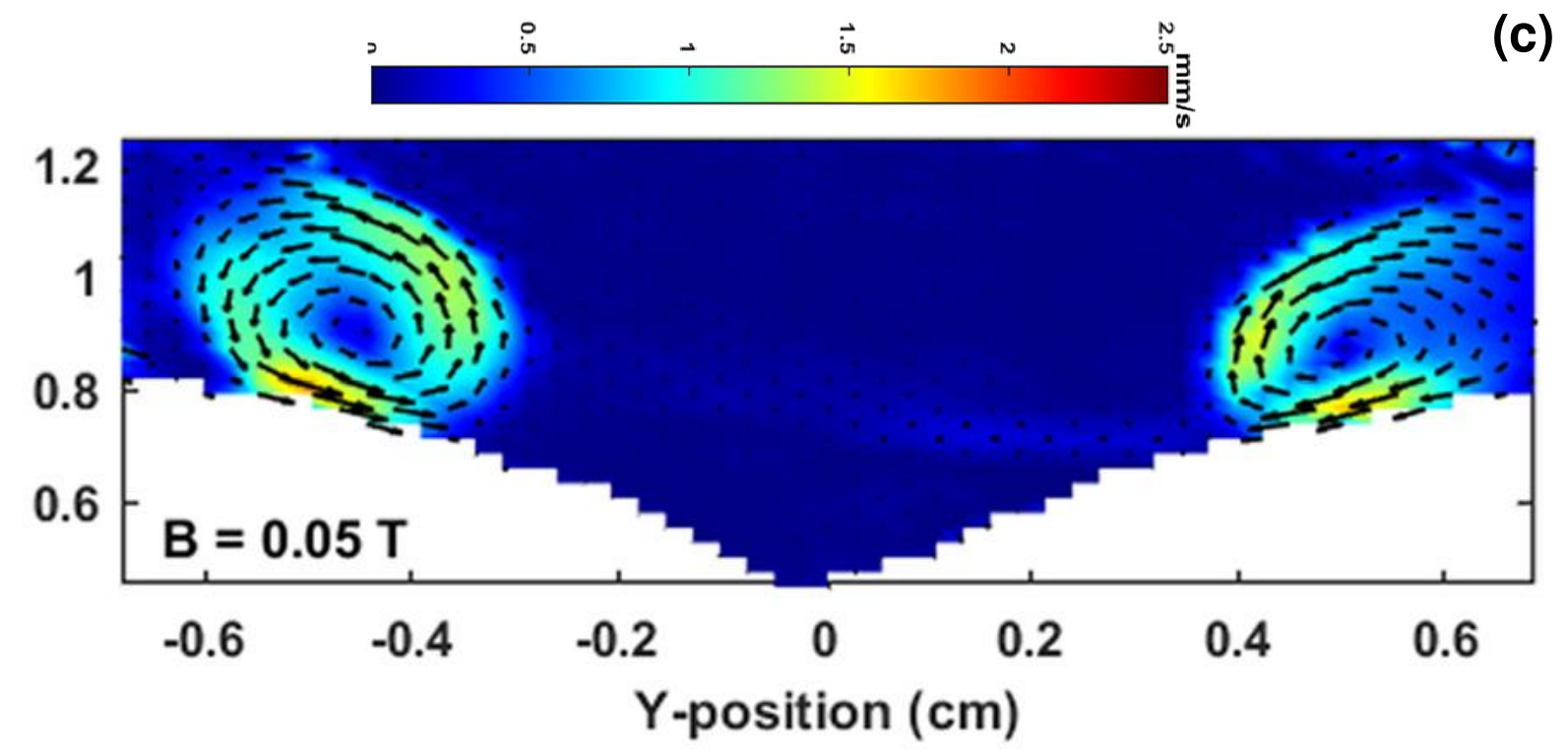}}}%
 \caption{\label{fig:fig7}(a) Eighteen overlapping video frames, side views, major ticks 2 mm. An RF voltage of 90 V (peak-to-peak) at pressure, 0.5 torr were fixed during experiments. The probe (wire) was biased at 30 V. Three images (i)--(iii) were taken at different heights (9mm, 8 mm, and 7 mm) of probe (white dot in image). These images in Fig.\ref{fig:fig7}(a) are reprinted with the permission from [Law \textit{et al.}, Phys. Rev. Lett. 80, 4189–4192 (1998)]  Copyright (1998)  by the American Physical Society. (b) A video image showing the vortex motion is reconstructed by the superimposition of five consecutive still images. An aluminum ring shaped electrode was used to confine dust particle. The yellow arrows represent the direction of vortex flow in this plane. The external magnetic field of strength 0.05 T was applied along the discharge axis. (c) PIV image of the corresponding video image of Fig.\ref{fig:fig7}(b)). This image is constructed after averaging the velocity vectors over 50 frames. The arrows indicate the direction of rotation in a vortex and the color bars represent the magnitude of the velocity of the rotating particles. Fig.\ref{fig:fig7}(b) and Fig.\ref{fig:fig7}(c) are reproduced from [Choudhary \textit{et al.}, Phys. Plasmas 27, 063701 (2020)], with the permission of AIP Publishing.}
 \end{figure*}

In daily life, we see the naturally occurring or induced vortices in fluids such as whirlpools in rivers and tornadoes. The vortices which are induced by any external force on a fluid element are termed forced vortices. In fluids, such vortices can be induced by rotating a vessel containing fluid or by paddling in a fluid. For studying the naturally occurring vortices, a full understanding of vortex behavior at the microscopic level is required. To demonstrate the vortex motion for graduate students in the physics laboratory, dusty plasma can be considered as a model system.
In a dust grain medium, vortex motion can be induced around a charged probe (metal wire) in unmagnetized RF discharge \cite{probeinducedcirculation} (see Fig.\ref{fig:fig7}(a)). An external magnetic field can also be used to excite the vortex motion in dust grain medium\cite{mangilalvortextorimagnetized}. Five consecutive still images are used to reconstruct the image of Fig.\ref{fig:fig7}(b) where one can observe the vortex motion in dusty plasma at a given B-field. The free available particle image velocimetry (PIV) code \cite{piv} and ImageJ software \cite{imagejsoftware} are very useful to obtain the velocity profile of vortex flow and angular velocity distribution of particles in a given vortex \cite{williamspivanalysis,mangilalvortextorimagnetized}.
A PIV image corresponding to the vortex flow in the presence of magnetic field \cite{mangilalvortextorimagnetized} is depicted in Fig.\ref{fig:fig7}(c). The dusty plasma device either in unmagnetized or magnetized RF discharge configuration can be used to demonstrate the vortex motion in fluids. Students can visualize the vortex motion at particle level in the dusty plasma medium and correlate these results to understand the vortices in other fluids. \par
\begin{figure*}
 \centering
\subfloat{{\includegraphics[scale=0.4200]{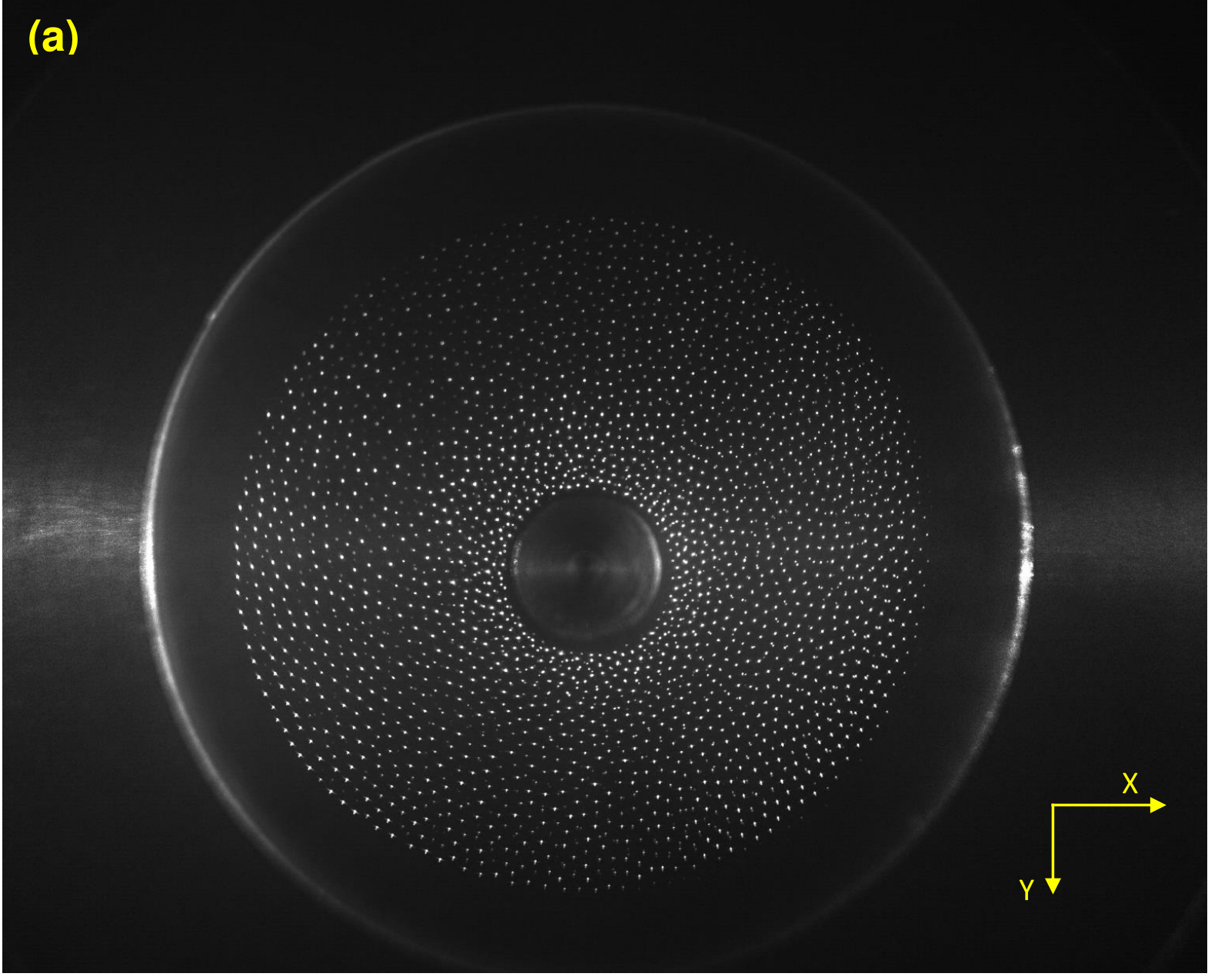}}}
\hspace*{0.35in}
 \subfloat{{\includegraphics[scale=0.7700]{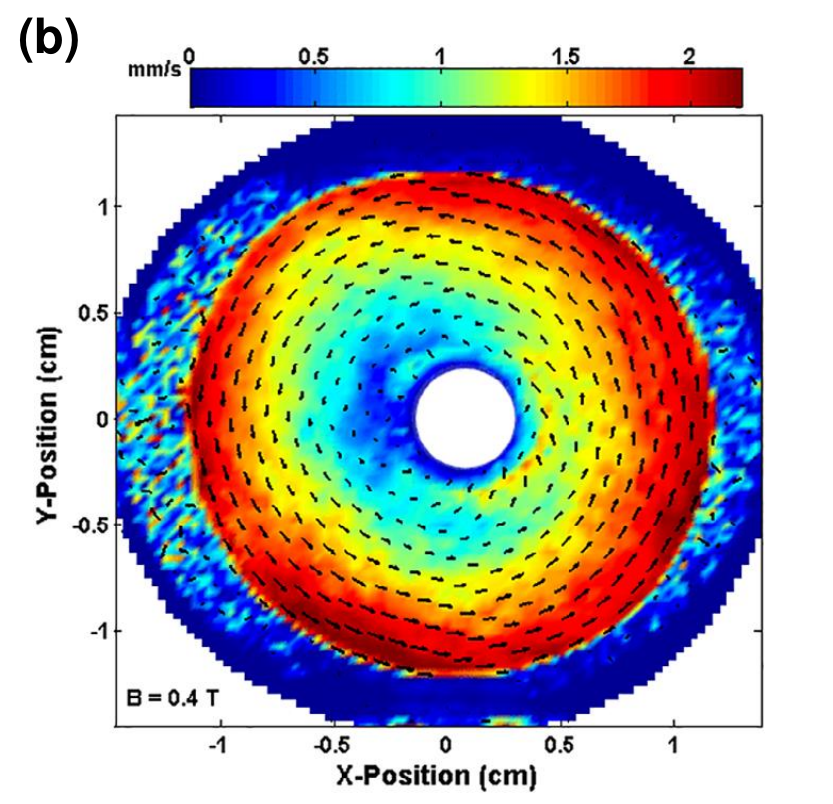}}}%
 \caption{\label{fig:fig8}(a) Dust grains confined in an annular region between a metal disk and ring. The dust grains are confined between an aluminium
disk of diameter 5 mm and ring diameter of 30 mm at electrodes voltage
$V_{up}$ = 55 V, $V_{down}$ = 55 V and argon pressure p = 30 Pa. White dots represent  charged dust particles \cite{mangilalannulusdusty}. (b) PIV image of the
rotational motion of dust grains in the X–Y plane at magnetic field of strength, B = 0.4 T (image ref.\cite{mangilalannulusdusty}).}
 \end{figure*}

Apart from the vortex motion, dust grain medium can also be used to demonstrate the rigid rotational motion of any medium. In the presence of a weak magnetic field (B $<$ 0.05 T), dust particles in either DC or RF discharge configuration exhibits the rigid rotational motion \cite{knopkamagneticrotation,mangilalannulusdusty}. It provides a platform to learn the rotational motion of many-body systems by estimating the angular frequency of rotating particles. Students may come to know the differences in translation and rotational motion. A single frame video image of an annulus dusty plasma is shown in Fig.\ref{fig:fig8}(a). The rigid body rotational motion of dust grains in the region of an annual in the presence of B-field is estimated by a PIV image (see Fig.\ref{fig:fig8}(b)). A constant angular frequency variation along radial direction indicates the rigid rotational motion of medium \cite{mangilalannulusdusty}. Thus, dusty plasma experiments at the graduate level may introduce students to the concepts of rotational motion of many-body system and vortex flow in fluids. 
\section{Image Processing and Analysis Techniques} \label{sec:image_analysis}
\begin{figure*}
 \centering
 \subfloat{{\includegraphics[scale=0.800]{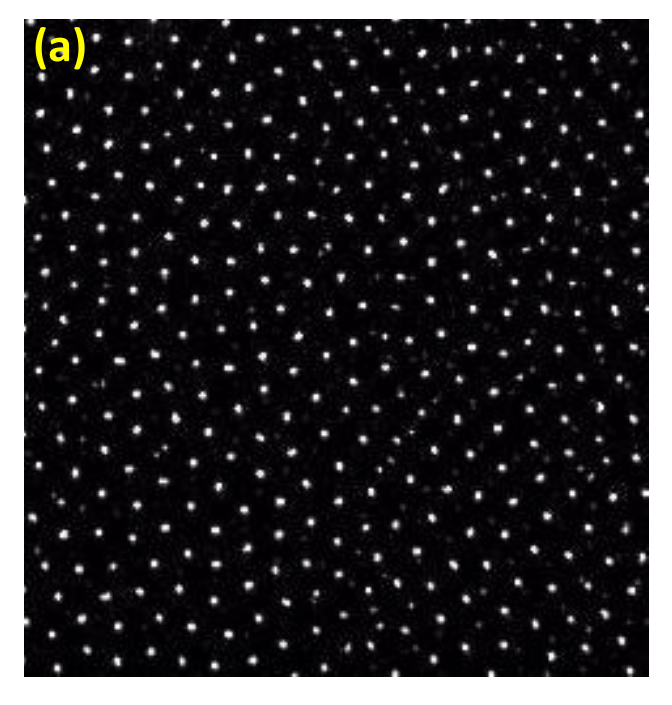}}}
\hspace*{0.25in}
 \qquad
 \subfloat{{\includegraphics[scale=0.800]{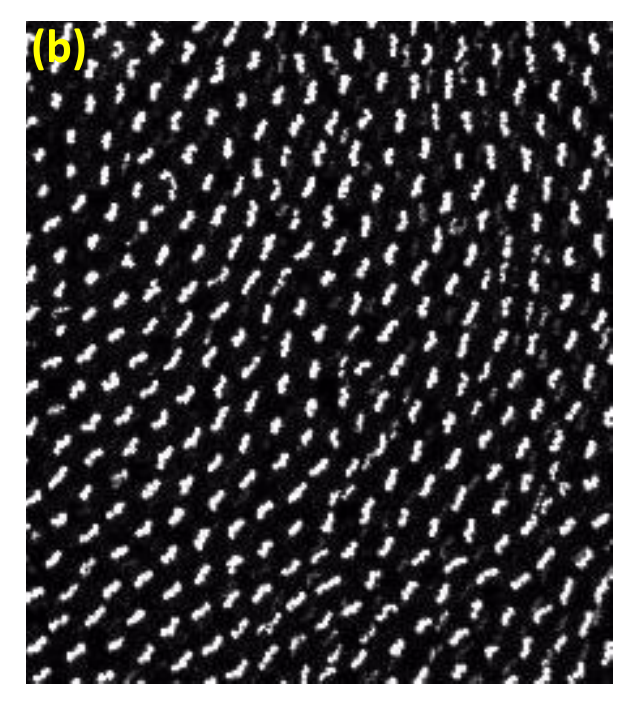}}}
 \caption{\label{fig:fig9}(a) A typical image of dust grain medium. (b) Image reconstructed by the superimposition of six consecutive still images correspond to  fig.\ref{fig:fig9}(a).}
 \end{figure*}

In dusty plasma experiments, the scattered lights from the solid charged particles are captured using a fast frame CCD or CMOS camera and image data (frames) are transferred to PC. These stored images are later analyzed to get the dynamics of dust grain medium for given discharge conditions. It provides a platform to undergraduate and post-graduate physics students for understanding the basics of image processing and image analysis techniques using various software and computational tools such as ImageJ, MATLAB, Python, etc. Using this dusty plasma device, students can get various kinds of dusty plasma data in form of images such as dust grain oscillation, dust-acoustic waves, rotational motion, linear flow of dust particles, dusty plasma crystal, etc. They can use these images to learn various tools and techniques for analyzing images and calculate the dusty plasma parameters which are helpful to understand the dynamics of dusty plasma. A typical raw image of dust grain medium and superimposition of six images (composite image) are shown in Fig.\ref{fig:fig9}. The superimposition of six images is done with the help of ImageJ software. In other examples, as shown in Fig.\ref{fig:fig3}(d) and Fig.\ref{fig:fig7}(c),  MATLAB image processing tools are used to get the intensity profile of DAWs at different times and velocity profile of dust particles respectively. Thus, such hands-on experience will help students gain a better understanding of the image processing and image analysis techniques using ImageJ, MATLAB, Python, and other software.
\begin{figure*}
 \centering
\subfloat{{\includegraphics[scale=0.800]{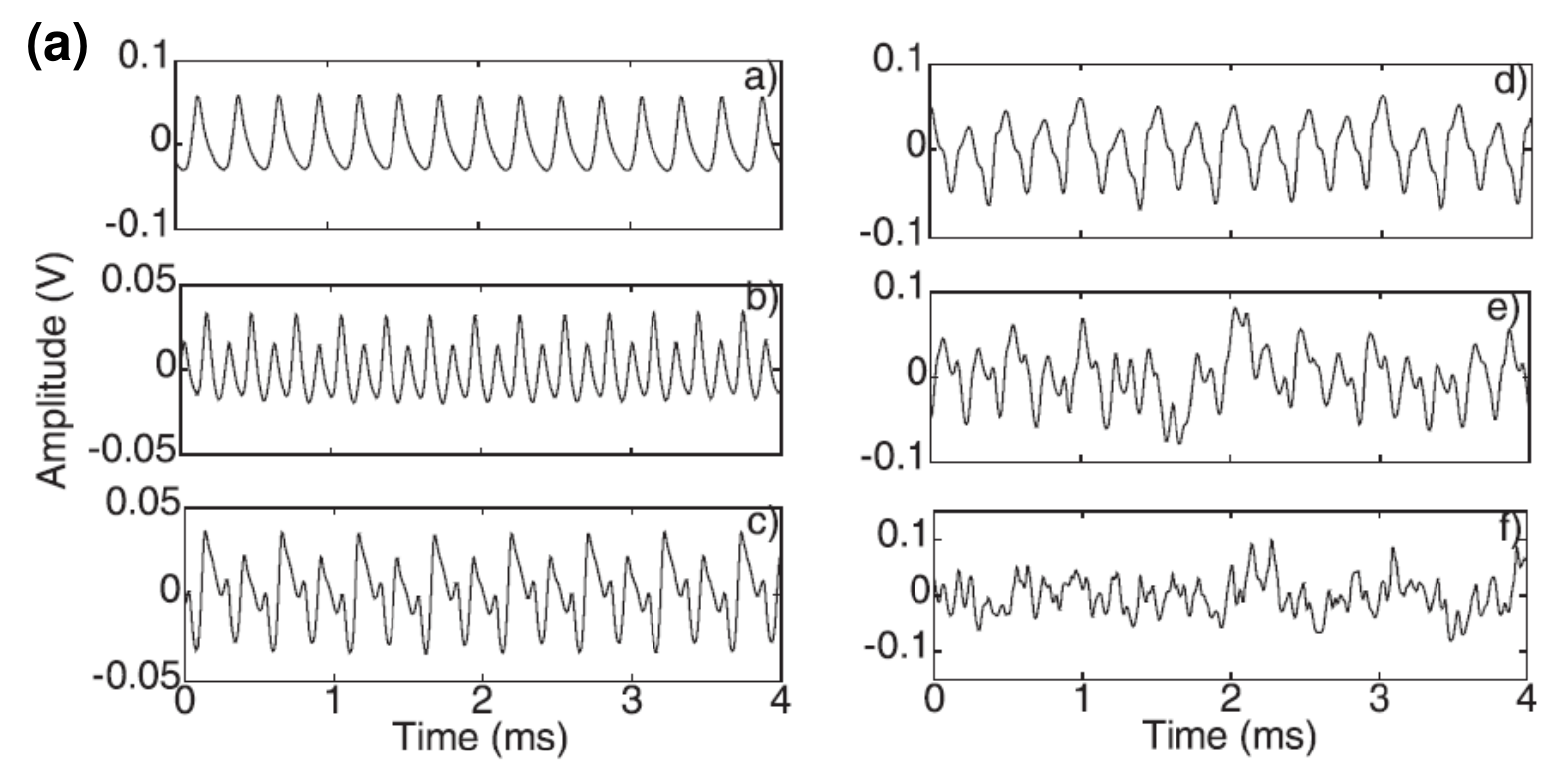}}}
 \qquad
 \subfloat{{\includegraphics[scale=0.800]{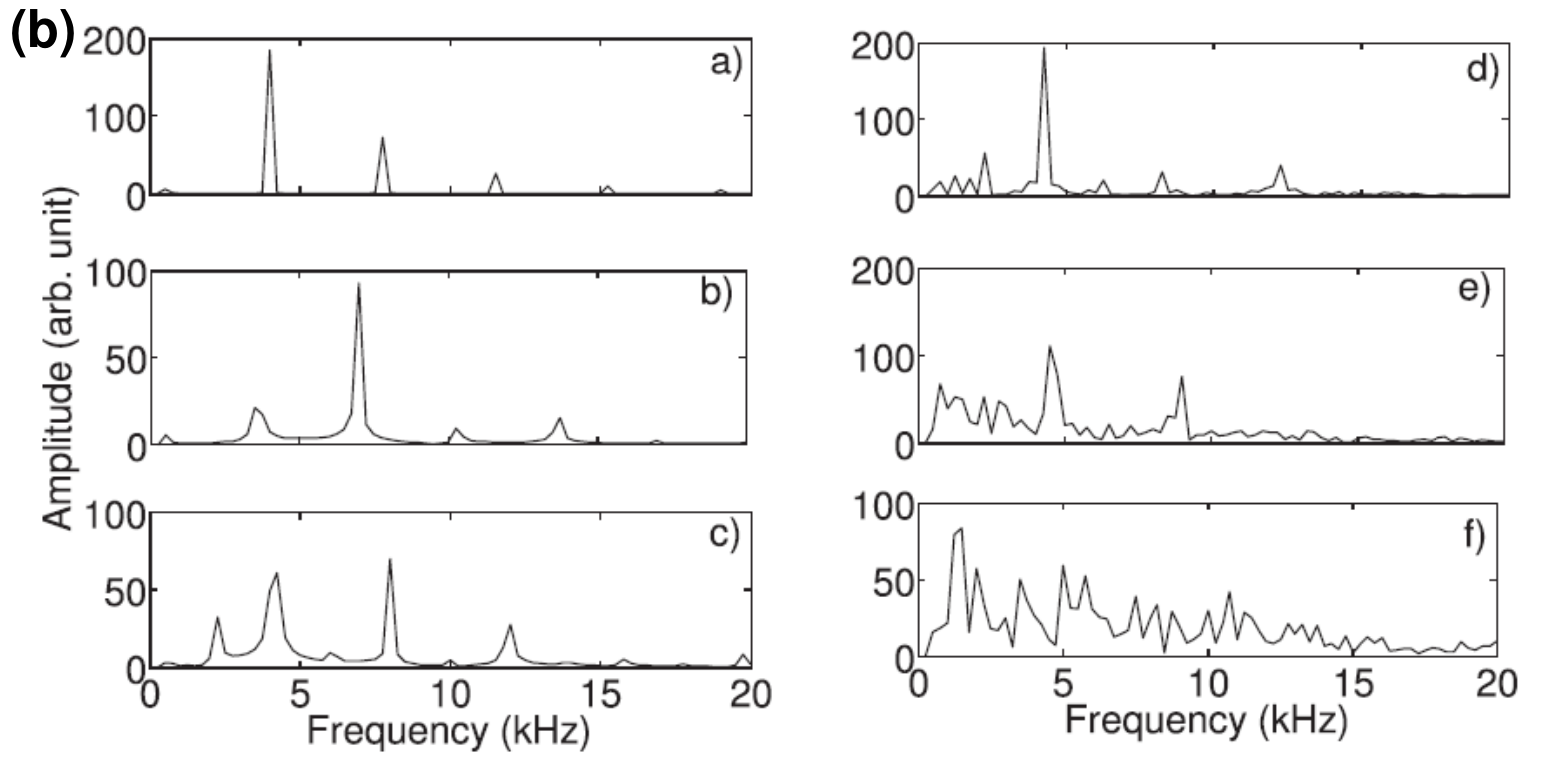}}}
 \qquad
\subfloat{{\includegraphics[scale=0.800]{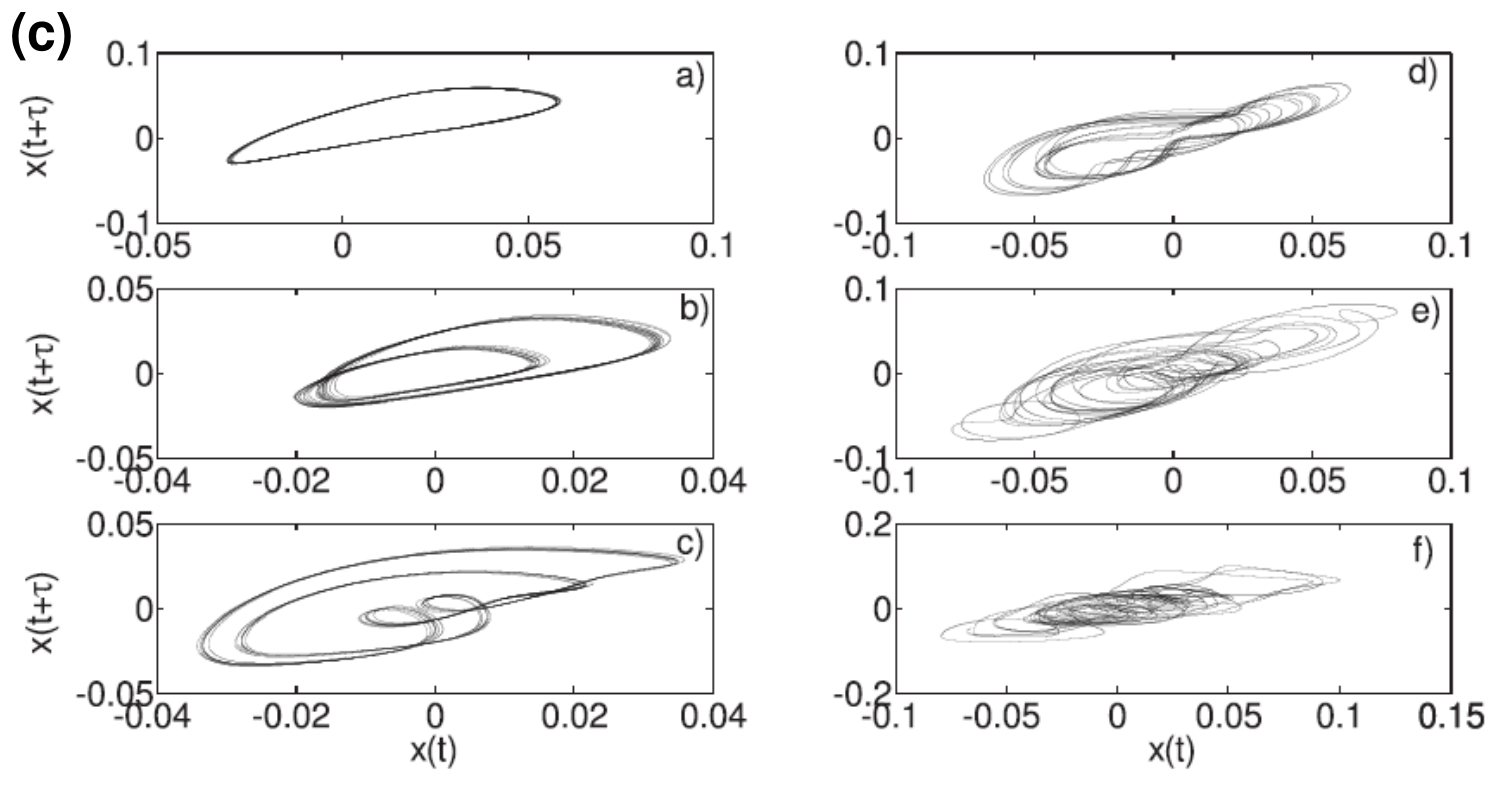}}}
 \caption{\label{fig:fig10}(a) Time series plots of ﬂoating potential ﬂuctuations at different values of magnetic field (fixed magnet).(b) Power spectrum plots corresponding to the time series data of Fig.\ref{fig:fig10}. (c) Reconstructed phase space plots corresponding to the time series data of Fig.\ref{fig:fig10}.
All Figures are reproduced from [Shaw \textit{et al.}, Phys. Plasmas 24, 082105 (2017)], with the permission of AIP Publishing.}
 \end{figure*}
\section{Time-series Data Analysis} \label{sec:data_analysis}
Students at the graduation level have knowledge of nonlinear (complex)  systems existing in the universe. Nonlinear systems exhibit sensitive dependence to the initial conditions of the systems. For example, a double pendulum undergoing large oscillations. It is difficult to predict the exact future trajectories of the oscillator (double pendulum) \cite{nonlinearsystemstrogate}. There are many natural physical nonlinear systems such as atmosphere, weather, climate change, wind speed, biological phenomena (blood pressure, heart rate), etc. The irregular temporal behavior of the variables for a given system (time-series data) is an output of such complex systems with nonlinear feedback loops and external driving force \cite{nonlinearsystemstrogate}. It is required to understand and model such irregular fluctuations to understand such highly nonlinear systems in the physical world. For understanding the nonlinear dynamical behavior of complex systems, various times-series data analysis methods such as fast Fourier transformation (FFT), phase space diagram, Lyapunov exponent, etc. are to be learned \cite{pankajiimeseries1,neerajnonliner1,pankajnonliner2}.\par
A DC glow discharge plasma is assumed to be a highly complex non-linear system. It can be used as a non-linear system to understand other natural or artificial highly nonlinear systems. The temporal irregular data (time-series data) in the form of discharge current or floating potential of plasma (with or without particles) are recorded at a given discharge condition \cite{pankajiimeseries1,neerajnonliner1,neerajnonliner2}. The pattern of the time-series data depends on the discharge parameters such as gas pressure, discharge voltage, external magnetic field, etc. 
In Fig.\ref{fig:fig10}(a), time-series data (floating potential) of a DC glow discharge plasma are displayed. The FFT of the same time series data is shown in Fig.\ref{fig:fig10}(b) from where one can obtain the value of frequency of fundamental mode as well as higher-order harmonics \cite{pankajiimeseries1}. The harmonics appear with an integer of the fundamental oscillation frequency which suggests the nonlinear behavior of the plasma medium. The phase space diagrams of the different temporal fluctuation are depicted in Fig.\ref{fig:fig10}(c). With changing the discharge parameters, the transition from a stable state to a chaotic state occurs. The transition from chaotic to periodic also possible with changing the discharge parameters. Apart from this nonlinear dynamical behavior, the plasma medium could also show the periodic oscillations and limit cycle oscillations that could be checked by different data analysis tools. Thus, DC glow discharge plasma provides a good platform for undergraduate and postgraduate students to understand the dynamical behaviour of highly nonlinear systems, predict the irregular fluctuations, and learn the time-series data analysis techniques.
\section{Summary}     \label{sec:summary}
In this perspective paper, the role of dusty plasma experiments in the learning process of undergraduate and post-graduate physics students at higher institutions/universities is discussed. I have proposed some basic dusty plasma experiments like waves and oscillations, diffraction of waves, crystallization, phase transition, vortex motion, rigid rotational motion, and data analysis techniques to demonstrate some basic physics experiment, create a scientific temper among graduate students, and provide a platform for learning experimental tools and techniques. How a single dusty plasma device either in direct current or radio-frequency discharge configuration can be used to perform various basic physics experiments as well as to learn various image and data analysis tools and techniques. A detailed discussion on each dusty plasma experiment and data analysis tools are presented in this paper. However, this paper only highlights opportunities for physics graduate students for performing some basic experiments in the physics lab using the dusty plasma device which can be operated either in DC or RF discharge configuration. The main focus of this article to highlight only the advantages of dusty plasma experiments to physics students by citing the previous experimental studies. A detailed procedure (or tutorial) for an individual experiment could be a future scope. 
 \section{Acknowledgement} 
 The author is grateful to Prof. Merlino, Prof. Lin I, Prof. Hyde and Dr. Shaw for allowing him to reuse the published figures with permission of publishers. Author is also thankful to Dr. R. Rajawat, Dr. V. Kella and Dr. A. Gupta for careful reading of this paper.
\bibliography{referencesdusty2020}
\end{document}